\documentclass[11pt]{article}
\usepackage{amssymb}
\usepackage{amsmath}
\usepackage{amscd}
\usepackage{latexsym,verbatim}
\usepackage[all]{xy}
\xyoption{curve}
\xyoption{line}
\usepackage{graphicx}
\usepackage{graphics}
\usepackage{color}
\usepackage[vcentermath,enableskew]{youngtab}
\catcode`@=11
\renewcommand{\section}
{\@startsection{section}{1}{0pt}{\medskipamount}{\medskipamount}{\large\bf}}
\makeatletter\renewcommand{\subsection}
{\@startsection{subsection}{2}{\z@}{-3.25ex plus -1ex minus -.2ex}
{1.5ex plus .2ex}{\it }}

\numberwithin{equation}{section}
\catcode`@=12

\def\b{\beta}

\def\m{\mu}
\def\n{\nu}
\def\pa{\partial}

\def\Dots{\cdot\cdot}
\def\mt{{\widetilde b}}
\def\indint{{b}}
\def\ct{{\widetilde c\,}}
\def\Dirac{{D\!\!\!\!/\,}} % Dirac operator
\def\Diraccal{{{\cal D}\!\!\!\!/\,}} % Dirac operator

\def\pa{{\partial\!\!\!/}}
\def\cp2star{\star}

\def\beq{\begin{equation}}
\def\eeq{\end{equation}}
\def\bea{\begin{eqnarray}}
\def\eea{\end{eqnarray}}

\newcommand{\im}{\,\mathrm{i}\,}
\newcommand{\diff}{\mathrm{d}}

\newcommand{\R}{{\mathbb{R}}}

\newcommand{\C}{{\mathbb{C}}}
\newcommand{\Z}{{\mathbb{Z}}}

\newcommand{\CP}{{\C P}}

\newcommand{\Idd}{\mathbf{1}}
\newcommand{\Hcal}{{\cal H}}
\newcommand{\Ecal}{{\cal E}}

\newcommand{\Vcal}{{\cal V}}
\newcommand{\Ical}{{\cal I}}

\newcommand{\Qcal}{{\cal Q}}

\newcommand{\Gcal}{{\cal G}}

\newcommand{\yb}{{\bar{y}}}
\newcommand{\Yb}{{\bar{Y}}}

\newcommand{\ab}{{\bar{a}}}

\newcommand{\jb}{{\bar{\jmath}}}
\newcommand{\ib}{{\bar{\imath}}}

\newcommand{\betab}{{\bar{\beta}}}

\newcommand{\ca}{{\cal{A}}}
\newcommand{\cf}{{\cal{F}}}

\newcommand{\man}{{\cal M}}
\newcommand{\cliff}{{{\rm C}\ell}}

\newcommand{\HQ}{{\rm H}}

\newcommand{\ch}{{\mathrm{ch}}}

\newcommand{\Lcal}{{\cal L}}

\newcommand{\Tr}{{\rm Tr}}

\newcommand{\tr}{{\rm tr}}

\newcommand{\su}{{{\rm SU}(2)}}
\newcommand{\suL}{{{\rm su}(2)}}
\newcommand{\sut}{{{\rm SU}(3)}}
\newcommand{\uo}{{{\rm U}(1)}}
\newcommand{\uoL}{{{\rm u}(1)}}
\newcommand{\uk}{{{\rm U}(p)}}

\newcommand{\urm}{{{\rm U}}}
\newcommand{\urmL}{{{\rm u}}}
\newcommand{\ut}{{{\rm U}(3)}}
\newcommand{\utwo}{{{\rm U}(2)}}

\newcommand{\utwoL}{{{\rm u}(2)}}
\newcommand{\Sp}{{\rm S}}
\newcommand{\SU}{{\rm SU}}

\newcommand{\spin}{{\rm Spin}}

\newcommand{\quiver}{{\sf W}}
\newcommand{\rel}{{\sf Q}}

\newcommand{\mphi}{{{\mbf\phi}}}

\newcommand{\mbf}[1]{{\boldsymbol {#1} }}

\newcommand{\xr}{\xrightarrow}
\newcommand{\noverq}{{\stackrel{\scriptstyle n}{\scriptstyle q}}}

\newcommand{\noverqpmt}{{\stackrel{\scriptstyle n}{\scriptstyle
      q\pm2}}}

\newcommand{\noverqpt}{{\stackrel{\scriptstyle n}{\scriptstyle q+2}}}
\newcommand{\noverqmt}{{\stackrel{\scriptstyle n}{\scriptstyle q-2}}}

\newcommand{\npoverqmo}{{\stackrel{\scriptstyle n+1}{\scriptstyle
      q-1}}} 
\newcommand{\nmoverqmo}{{\stackrel{\scriptstyle n-1}{\scriptstyle
      q-1}}} 
\newcommand{\npoverqpo}{{\stackrel{\scriptstyle n+1}{\scriptstyle
      q+1}}} 
\newcommand{\nmoverqpo}{{\stackrel{\scriptstyle n-1}{\scriptstyle
      q+1}}} 
\newcommand{\npmoverqpo}{{\stackrel{\scriptstyle n\pm1}{\scriptstyle
      q+1}}} 
\newcommand{\npmoverqmo}{{\stackrel{\scriptstyle n\pm1}{\scriptstyle
      q-1}}}

\def\Hom{{\rm Hom}}

\def\End{{\rm End}}

\def\>{\rangle}
\def\<{\langle}
\def\+{\dagger}
\def\={\ =\ }

\begin{document}

\begin{titlepage}
\setcounter{page}{0}
\begin{flushright}
DIAS--STP--09--06\\
pi--partphys--133\\
HWM--09--4\\
EMPG--09--7\\
\end{flushright}

\vskip 1.8cm

\begin{center}

{\Large\bf Dimensional Reduction and \\[10pt]
Vacuum Structure of Quiver Gauge Theory}

\vspace{15mm}

{\large Brian P. Dolan${}^{1,2}$}
\ \ and \ \ {\large Richard J. Szabo${}^{3,4}$}
\\[5mm]
\noindent ${}^1${\em Department of Mathematical Physics, National
  University of Ireland\\ Maynooth, Co. Kildare, Ireland}
\\[5mm]
\noindent ${}^2${\em Perimeter Institute for Theoretical Physics\\ 31
  Caroline St. N, Waterloo, Ontario N2L 2Y5, Canada}
\\[5mm]
\noindent ${}^3${\em Department of Mathematics, Heriot-Watt
  University\\ Colin Maclaurin Building, Riccarton, Edinburgh EH14
  4AS, U.K.}
\\[5mm]
\noindent ${}^4${\em Maxwell Institute for Mathematical Sciences\\
  Edinburgh, U.K.}
\\[5mm]
{Email: {\tt bdolan@thphys.nuim.ie , R.J.Szabo@ma.hw.ac.uk}}

\vspace{15mm}

\begin{abstract}
\noindent
We describe the structure of the vacuum states of quiver gauge
theories obtained via dimensional reduction over homogeneous spaces,
in the explicit example of $\sut$-equivariant dimensional reduction of
Yang-Mills-Dirac theory on manifolds of the form $M\times\CP^2$. We
pay particular attention to the role of topology of background gauge
fields on the internal coset spaces, in this case $\uo$ magnetic
monopoles and $\su$ instantons on $\CP^2$. The reduction of Yang-Mills
theory induces a quiver gauge theory involving coupled
Yang-Mills-Higgs systems on $M$ with a Higgs potential leading to
dynamical symmetry breaking. The criterion for a ground state of the
Higgs potential can be written as the vanishing of a non-abelian
Yang-Mills flux on the quiver diagram, regarded as a lattice with
group elements attached to the links. The reduction of
$\sut$-symmetric fermions yields Dirac fermions on $M$ transforming
under the low-energy gauge group with Yukawa couplings. The fermionic
zero modes on $\CP^2$ yield exactly massless chiral fermions on $M$,
though there is a unique choice of spin$^c$ structure on $\CP^2$ for
which some of the zero modes can acquire masses through Yukawa
interactions. We work out the spontaneous symmetry breaking patterns
and determine the complete physical particle spectrum in a number of
explicit examples, some of which possess quantum number assignments
qualitatively analogous to the manner in which vector bosons, quarks
and leptons acquire masses in the standard model. 
\end{abstract}
\end{center}
\end{titlepage}
 
%\tableofcontents

\newpage

\section{Introduction\label{Intro}}

\noindent
The Kaluza-Klein programme, i.e. the idea that the Higgs and Yukawa
sectors of the standard model of particle physics could have their
origins in a simpler but higher-dimensional theory, remains as
attractive today as when it was when first proposed~\cite{KK}. In the
original idea of Kaluza and Klein, and its non-abelian generalisation
with a homogeneous internal space $G/H$ for $H$ a closed subgroup of a
compact Lie group $G$, the higher-dimensional theory was pure gravity
but in later schemes Einstein-Yang-Mills theories in higher dimensions
were introduced~\cite{KKReviews}.  This has the potential to provide a
unification of the gauge and Higgs sectors in higher dimensions, while
the coupling of fermions to the higher-dimensional gauge theory
naturally induces Yukawa couplings after dimensional reduction. For
certain coset spaces, particularly the complex projective plane, the
inclusion of topologically non-trivial internal fluxes can induce the
chiral fermionic spectrum of quarks and leptons of the standard
model~\cite{BDCN}.

The pioneering scheme realizing these constructions is called ``coset
space dimensional reduction''~\cite{FMT,KZ1}, though a generic problem
with such reductions has been that they are unable to generate
chiral gauge theories, without some additional
modifications~\cite{KZ1,ShelterIsland}. In coset space dimensional
reduction, constraints are imposed on the higher-dimensional fields
ensuring that they are invariant under the $G$-action up to gauge
transformations. This amounts to studying embeddings of the isometry
group $G$ of the coset space, or of its holonomy subgroup $H$, in the
gauge group of the higher-dimensional theory and the solutions of the
constraints are then formally identified with the lowest modes of the
harmonic towers of fields.

On the other hand, the ``equivariant dimensional reduction'' of gauge
theories naturally incorporates the topology of background fields on
$G/H$ which are gauged with respect to the holonomy group
$H$. Although similar in spirit to the coset space dimensional
reduction scheme, equivariant dimensional reduction systematically
constructs the unique field configurations on the higher-dimensional
space which are equivariant with respect to the internal isometry
group $G$, and reduces Yang-Mills theory to a quiver gauge theory
based on a quiver (with relations) which is determined entirely by the
representation theory of the Lie groups $G$ and $H$. As
in coset space dimensional reduction, there is no \emph{a priori}
relation between the gauge group $\Gcal$ of the higher-dimensional
field theory and the groups $G$ or $H$, and the resulting gauge group
of the dimensionally reduced field theory is a subgroup of
$\Gcal$. This differs from the usual Kaluza-Klein reductions where the
isometry group (or the holonomy group) is identified with the gauge
group. The general formalism is described in~\cite{ACGP1,LPS1}. It has
been applied in a variety of contexts in~\cite{LPS2,Dolan09} when the
internal coset space is the projective line $\CP^1$. Dimensional
reduction over the fuzzy sphere $\CP_F^1$ is also considered
in~\cite{Dolan09,ASMMZ}. In this paper we extend the analysis of the
vacuum states of such quiver gauge theories performed
in~\cite{Dolan09} to an example with non-abelian holonomy, the
projective plane $\CP^2$. The corresponding quiver gauge theories
have been discussed in~\cite{LPS3}. This example is rich enough to
capture some general features of the vacua of the quiver gauge
theories which are induced by reduction over generic coset spaces
$G/H$.

When the internal space is the projective plane $\CP^2$, the
equivariant dimensional reduction of gauge fields naturally comes with
$\uo$ monopoles and $\su$ instantons, in contrast to $\CP^1$ where
only monopoles are present, and this introduces essential differences
from the $\CP^1$ case. As in the Kaluza-Klein approach, the mass scale
of the dimensionally reduced field theory is set by the size of the
internal space. We obtain a Higgs sector of the lower-dimensional
gauge theory with a Higgs potential that leads to dynamical symmetry
breaking, as a direct consequence of the non-trivial internal fluxes,
and we work out the complete physical particle content and
masses for a number of explicit symmetry breaking hierarchies. As in
the case of reduction over $\CP^1$, a Yukawa sector of the reduced
fermionic field theory is naturally induced. The harmonic expansion
over $\CP^2$ induces an infinite tower of massive fermions in the
reduced field theory, but the topologically non-trivial gauge fields
on the internal $\CP^2$ necessarily also induce exactly massless
chiral modes in the reduced field theory. As in the $\CP^1$ case, some
of the massless spinor fields which arise as a consequence of the
index theorem on the internal space acquire masses through their
Yukawa couplings, but in general not all of them.

There is a number of other differences between the equivariant
dimensional reduction over $\CP^1$ and that over $\CP^2$ which is
studied here. The fact that the rank of the holonomy group is now
greater than one means that the quiver diagram is no longer a
one-dimensional chain but is a higher-dimensional lattice, of
dimension two in the case of $\CP^2$. We show that the condition for a
vacuum state of the Higgs sector of the reduced field theory can be
phrased in terms of a non-abelian gauge theory on the quiver
lattice. A group element associated with the Higgs field can be placed
on each link of the quiver diagram, and minimising the Higgs potential
requires that the resulting gauge field flux on the quiver lattice is
zero. The Higgs vacuum requires that the lattice gauge field is gauge
equivalent to the trivial gauge potential. 

Another difference is associated with spinors on $\CP^2$ and the
treatment of the fermionic field theory. It is well-known that $\CP^2$
does not admit a spin structure, as there is a global obstruction to
putting spinors on $\CP^2$ associated with the fact that its second
Stiefel-Whitney class is non-trivial~\cite{Milnor}. However, since the
equivariant dimensional reduction scheme necessarily induces
topologically non-trivial monopole and instanton fields on the
internal space, the reduction itself provides a solution to the
problem of absence of spin structure on $\CP^2$ by simply coupling
spinor fields to non-trivial gauge backgrounds and using spin$^c$
structures for line bundles or non-abelian spin$^c$ structures for
higher rank bundles. Gauge fields on $\CP^2$ and their coupling to
spinors were studied in \cite{BDCP2,CNash}. We will find that there is
a unique spin$^c$ structure accommodating the background
gauge fields on $\CP^2$ which generically lead to Yukawa interactions
after dimensional reduction, in contrast to the $\CP^1$ reductions,
whereas other choices of twisting can produce more realistic
generations of fermions. Altogether, we will explicitly display models
in which the quantum number assignments for the fermions are
qualitatively similar to those of quarks and leptons in the standard
model. 

This paper is organised as follows. In \S\ref{dimred} we describe the
kinematics of equivariant dimensional reduction over $\CP^2$,
particularly how the gauge and Higgs fields in the reduced field
theory depend on representation theory and the various irreducible
$\SU(2)\times\urm(1)$ representations that can arise from a given
$\SU(3)$ representation, as well as the harmonic expansion of zero
mode spinors. In \S\ref{Eqgaugeth} we derive the dimensionally reduced
action, showing how the Higgs potential depends on the group
representation content and how Yukawa couplings are induced in the
Dirac action. Some examples are studied in detail in the ensuing two
sections, one class of examples based on the fundamental
representation of $\sut$ in \S\ref{Fundamental} and one class based on
the adjoint representation in \S\ref{Adjoint}. Our conclusions are
summarised in \S\ref{Conclusions}. Some technical details are
relegated to three appendices at the end of the paper. In
\S\ref{ChernCharacter} we calculate Chern numbers for the various
equivariant vector bundles over $\CP^2$ required in our analysis. Some
useful identities for equivariant one-forms on $\CP^2$ are given in
\S\ref{betamatrixprods}. Finally, the index of the Dirac operator on
$\CP^2$, coupled to various topologically non-trivial gauge field
backgrounds, is computed in \S\ref{ASItheorem}.

\bigskip

\section{Equivariant dimensional reduction over the projective 
  plane\label{dimred}}

\noindent
In this section we will describe the $\sut$-equivariant dimensional
reduction of gauge and fermion fields over an internal complex
projective plane $\C P^2$. For some further details,
see~\cite{LPS3}. Throughout this section 
all local coordinates and fields are taken to be dimensionless.

\subsection{Homogeneous vector bundles on $\C
  P^2$\label{HombunCP2}}

We are interested in the geometry of the symmetric coset space $\C
P^2\cong G/H$, where
\beq
H\=\Sp\big(\utwo\times\uo\big)~\cong~\su\times\uo
\label{holgp}\eeq
is the holonomy subgroup of the isometry group $G=\sut$ of $\C
P^2$. Given a finite-dimensional representation $\underline{V}$ of
$H$, the corresponding induced, homogeneous hermitean vector bundle
over $\C P^2$ is given by the fibred product
\beq
\mathcal{V}=G\times_H\,\underline{V} \ .
\label{indhermbungen}\eeq
Every $G$-equivariant bundle of finite rank over $\C P^2$, with
respect to the standard left transitive action of $G$ on the
homogeneous space, is of the form (\ref{indhermbungen}). If
$\underline{V}$ is irreducible, then $H$ is the structure group of the
associated principal bundle. We restrict to those representations
$\underline{V}$ which descend from some irreducible representation of
$\sut$ by restriction to $H$.

The Dynkin diagram for $\sut$ consists of a pair of roots
$\alpha_1,\alpha_2$. The complete set $\Delta$ of non-null roots is
$\pm\,\alpha_1,\pm\,\alpha_2,\pm\,(\alpha_1+\alpha_2)$, with the inner
products $(\alpha_1,\alpha_1)=(\alpha_2,\alpha_2)=1$ and
$(\alpha_1,\alpha_2)=-\frac12$ so that
$(\alpha_1+\alpha_2,\alpha_1+\alpha_2)=1$. For the system
$\Delta^+$ of positive roots we
take $\alpha_1=(1,0)$, $\alpha_2=\frac12\,(-1,\sqrt3\,)$ and
$\alpha_1+\alpha_2=\frac12\,(1,\sqrt3\,)$. The generators of $\sut$
for the Cartan-Weyl basis are given by the Chevalley generators
$E_{\alpha_1}$, $E_{\alpha_2}$ and
$E_{\alpha_1+\alpha_2}:=[E_{\alpha_1},E_{\alpha_2}]$, together with
the generators $H_{\alpha_1}$ and $H_{\alpha_2}$ of the Cartan
subalgebra $\uoL\oplus\uoL$. The non-vanishing commutation relations
are
\bea
[H_{\alpha_1},E_{\pm\,\alpha_1}]\=\pm\,2\,E_{\pm\,\alpha_1} \qquad
&\mbox{and}& \qquad [H_{\alpha_2},E_{\pm\,\alpha_1}]\=0 \ , \nonumber
\\[4pt] [H_{\alpha_1},E_{\pm\,\alpha_2}]\=
\mp\,E_{\pm\,\alpha_2} \qquad &\mbox{and}& \qquad
[H_{\alpha_2},E_{\pm\,\alpha_2}]\=\pm\,3\,
E_{\pm\,\alpha_2} \ , \nonumber\\[4pt]
[H_{\alpha_1},E_{\pm\,(\alpha_1+\alpha_2)}]\=\pm\,
E_{\pm\,(\alpha_1+\alpha_2)} \qquad &\mbox{and}& \qquad [H_{\alpha_2},
E_{\pm\,(\alpha_1+\alpha_2)}]
\=\pm\,3\,E_{\pm\,(\alpha_1+\alpha_2)} \ , \nonumber\\[4pt]
[E_{\alpha_1},E_{-\alpha_1}]\=H_{\alpha_1} \qquad &\mbox{and}& \qquad
[E_{\alpha_2},E_{-\alpha_2}]\=\mbox{$\frac12$}\,(H_{\alpha_2}-
H_{\alpha_1}) \ , \nonumber\\[4pt]
[E_{\alpha_1+\alpha_2},E_{-\alpha_1-\alpha_2}]\=\mbox{$\frac12$}\,
(H_{\alpha_1}+H_{\alpha_2}) \qquad &\mbox{and}& \qquad
[E_{\pm\,\alpha_1},E_{\pm\,\alpha_2}]\=E_{\pm\,(\alpha_1+\alpha_2)} \
, \nonumber\\[4pt]
[E_{\pm\,\alpha_1},E_{\mp\,(\alpha_1+\alpha_2)}]\=\mp\,
E_{\mp\,\alpha_2} \qquad &\mbox{and}& \qquad [E_{\pm\,\alpha_2},
E_{\mp\,(\alpha_1+\alpha_2)}]\=\pm\,E_{\mp\,\alpha_1} \ .
\label{HEEcomms}\eea

The fundamental weights are
$\mu_{\alpha_1}=\frac12\,\big(1,\frac1{\sqrt3}\,\big)$ and
$\mu_{\alpha_2}=\big(0,\frac1{\sqrt3}\,\big)$. For each pair of
non-negative integers $(k,l)$ there is an irreducible representation
$\underline{C}^{k,l}$ of $\sut$ of dimension 
\beq
d^{k,l}~:=~\dim\big(\,\underline{C}^{k,l}\,\big)
\=\mbox{$\frac12$}\,(k+1)\,(l+1)\,(k+l+2)
\label{dimkl}\eeq
and highest weight $\mu=k\,\mu_{\alpha_1}+l\,\mu_{\alpha_2}$. We label
the weight vectors of $\utwo\cong\su\times\uo$ in $\sut$ by
$(n,m)$ with respect to the basis $(H_{\alpha_1},H_{\alpha_2})$.
The eigenvalue of $H_{\alpha_1}$ is $n=2I$ and labels twice the
isospin $I$, so that $(n+1)$ is the dimension of the irreducible $\su$ 
representation. The eigenvalue $m=3Y$ of $H_{\alpha_2}$ is 
three times the hypercharge $Y$, and later on we shall identify $m$ 
with twice the magnetic charge.
The restriction of the $\sut$ operators $E_{\pm\,\alpha_1}$ to $\su$
shifts vertices along the horizontal directions of the weight
diagrams, while the generators $E_{\alpha_2}$ and
$E_{\alpha_1+\alpha_2}$ act on the weights as
\beq
(n,m)~\longmapsto~(n\pm1,m+3) \ ,
\label{CP2arrows}\eeq
depending on which particular weight vectors $(n,m)$ the raising
operators $E_{\alpha_1+\alpha_2}$ and $E_{\alpha_2}$ act on. 

For a fixed pair of non-negative integers $(k,l)$, the decomposition
of the irreducible $\sut$-module $\underline{C}^{k,l}$ as a
representation of $\su\times\uo$ can be obtained by collapsing the
``horizontal'' $\su$ representations to single nodes in the weight
diagram for $\underline{C}^{k,l}$. The corresponding collection of
weights $(n,m)$, which we denote by $\quiver_{k,l}$, is conveniently
parameterized by a pair of independent $\su$ spins $j_\pm=j_\pm(n,m)$,
with $2j_+=0,1,\dots,k$ and $2j_-=0,1,\dots,l$, that are defined in
terms of Young tableaux as follows. Represent the irreducible
$H$-module $\underline{(n,m)}$ with $(n,m)=(1,1)$ by
$\vcenter{\hbox{\tiny$\young(\times)$}}$ and that with $(n,m)=(0,-2)$
by $\vcenter{\hbox{\tiny$\young(\circ)$}}\,$. Then the
$\sut\rightarrow\su\times\uo$ decomposition of the fundamental
representation
\beq
\underline{C}^{1,0}\big|_H=\underline{(1,1)}\,\oplus\,
\underline{(0,-2)}
\label{fundrepdecomp}\eeq
is depicted by
\beq
\young(\ ) \qquad \longrightarrow \qquad \young(\times)\raise -5pt
\hbox{$\,\scriptstyle 1       $}\quad  \oplus
\quad \young(\circ)\raise -5pt \hbox{$\;\scriptstyle -2$} \ . 
\label{3to2tableau}
\eeq
In terms of $\SU(3)$ Young tableaux, the irreducible representation
$\underline C^{k,l}$ corresponds to the diagram
\begin{equation}
\underbrace{\young(\ \Dots\ ,\  \Dots \ )}_l\kern -2.1pt
\raise 6.2pt\hbox{$\underbrace{\young(\ \Dots\ )}_k$} \ ,
\end{equation}
and this contains all $\SU(2)\times \uo$ representations
\beq
\underbrace{\young(\times \Dots\times ,\times  \Dots \times
  )}_{l-2j_-}\kern -2.1pt\underbrace{\young(\circ \Dots\circ ,\times
  \Dots \times )}_{2j_-}\kern -2.1pt 
\raise 6.1pt\hbox{$\underbrace{\young(\circ
    \Dots\circ)}_{k-2j_+}$}\kern -0.5pt\raise
6.1pt\hbox{$\underbrace{\young(\times \Dots\times )}_{2j_+}$} 
\eeq
of dimension $2j_++2j_-+1$ and charge $2(l-k)+6(j_+-j_-)$, with
multiplicity one. This gives
\beq
n\=2(j_++j_-) \qquad \mbox{and} \qquad m\=6(j_+-j_-)-2(k-l) \ .
\label{mnqjpm}\eeq
The integers $(n,m)$ have the same even/odd parity. This is because
the weights come from embedding $\su\times\uo$ in $\sut$, and as such 
they only give faithful representations of $\urm(2)$. 

The bundle (\ref{indhermbungen}) with
$\underline{V}=\underline{C}^{k,l}\big|_H$ corresponds to a
representation of a certain finite quiver with relations~\cite{LPS3} 
in the category of homogeneous vector bundles over $\CP^2$. The
elements $(n,m)$ of the set $\quiver_{k,l}$ can be associated with
vertices of a directed graph
\beq
\mbox{\includegraphics[width=3in]{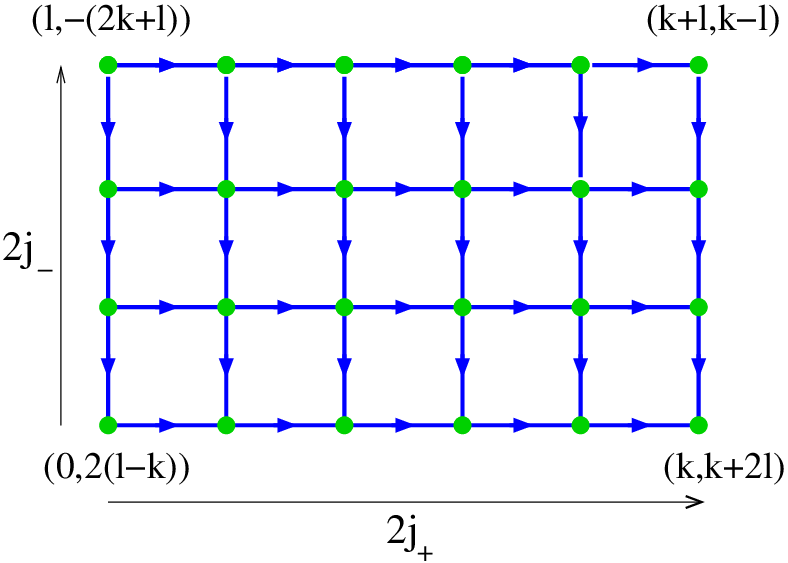}}
\label{Figquiver}\eeq
where only the four boundary corners are labelled with their values of
$(n,m)$ to avoid cluttering the diagram. The weight morphisms
(\ref{CP2arrows}) take the simple forms
\bea
j_+(n+1,m+3)\=j_+(n,m)+\mbox{$\frac12$} \qquad & \mbox{and} &
\qquad j_+(n-1,m+3)\=j_+(n,m) \ , \nonumber\\[4pt]
j_-(n-1,m+3)\=j_-(n,m)-\mbox{$\frac12$} \qquad & \mbox{and} &
\qquad j_-(n+1,m+3)\=j_-(n,m) \ ,
\label{SU2spinshifts}\eea
corresponding to the horizontal and vertical arrows in
(\ref{Figquiver}). We will refer to this graph as the ``quiver
lattice'', since the vacua of the quiver gauge theories we
consider later on have an elegant interpretation in terms of lattice
gauge theory defined on the directed graph (\ref{Figquiver}).

\subsection{$\sut$-equivariant bundles\label{fundquiverreps}}

We are interested in the structure of $G$-equivariant gauge fields
on manifolds of the form
\beq
\man~:=~M\times \C P^2\=G\times_HM \ ,
\label{M2ntimesSU3H}\eeq
where $M$ is a manifold of (real) dimension $d$ and $G=\sut$ acts
trivially on $M$. We will reduce gauge theory on
(\ref{M2ntimesSU3H}) by compensating the isometries of $\C P^2$ with
gauge transformations, such that the Lie derivative with respect to a
Killing vector field is given by an infinitesimal gauge transformation
on $\man$. This twisted reduction is accomplished by uniquely
extending the homogeneous vector bundles (\ref{indhermbungen}) by
$H$-equivariant bundles $E\to M$.

Let $\Ecal^{k,l}\to \man$ be a rank~$p$ hermitean vector bundle over
the space (\ref{M2ntimesSU3H}), associated to an irreducible
representation $\underline{C}^{k,l}$ of $\sut$, with structure group
$\urm(p)$. There is a one-to-one correspondence between
$G$-equivariant hermitean vector bundles over $\man$ and
$H$-equivariant hermitean vector bundles over $M$, with $H$ acting
trivially on $M$~\cite{ACGP1}. Given an $H$-equivariant bundle
$E^{k,l}\to M$ of rank $p$ associated to the representation
$\underline{C}^{k,l}\big|_H$ of $H$, the corresponding $G$-equivariant
bundle over $\man$ is defined by induction as
\beq
\Ecal^{k,l}=G\times_HE^{k,l} \ .
\label{Ecalklind}\eeq
The action of the holonomy group $H$ on $E^{k,l}$ is defined by the
isotopical decomposition
\beq
E^{k,l}\=\bigoplus_{(n,m)\in\quiver_{k,l}}\,E_{n,m}
\otimes\, \underline{(n,m)}\qquad \mbox{with} \quad
E_{n,m}\=\Hom_H\big(\,\underline{(n,m)}\,,\,E^{k,l}\,
\big) \ , 
\label{Eklisotopical}\eeq
where $\underline{(n,m)}$ are the irreducible $H$-modules occurring in
the decomposition of $\underline{C}^{k,l}\big|_H$. The vector bundles
$E_{n,m}\to M$ have rank $p_{n,m}$ and trivial $H$-actions. The
rank $p$ of $E^{k,l}$ is given by
\beq
p\=\sum_{(n,m)\in\quiver_{k,l}}\,(n+1)\,p_{n,m} \ .
\label{rankdimv}\eeq

The action of the $\sut$ operators $E_{\pm\,\alpha_2}$ and
$E_{\pm\,(\alpha_1+\alpha_2)}$ is implemented by means of
bi-fundamental Higgs fields
$\phi^\pm_{n,m}\in\Hom(E_{n,m},E_{{n\pm1,m+3}})$. These
bundle morphisms realize the $G$-action of the coset
generators which twists the naive dimensional reduction by
``off-diagonal'' terms. This construction explicitly breaks the gauge
group of the bundle $E^{k,l}$ as
\beq
\urm(p)~\longrightarrow~\prod_{(n,m)\in\quiver_{k,l}}\,\urm(p_{n,m})
\ .
\label{gaugegroupbreak}\eeq
With 
\beq
\Hcal_{n,m}=G\times_H\,\underline{(n,m)}
\label{Hcalnmdef}\eeq
the homogeneous bundle (\ref{indhermbungen}) induced by the
irreducible $H$-module $\underline{(n,m)}$, the structure group of the
principal bundle associated to
\beq
\Ecal^{k,l}=\bigoplus_{(n,m)\in\quiver_{k,l}}\,E_{n,m}
\boxtimes\Hcal_{n,m}
\label{Ecalklisotopical}\eeq
is then $H\times\prod_{(n,m)\in\quiver_{k,l}}\,\urm(p_{n,m})$.

\subsection{Canonical connections on $\C P^2$\label{canconnCP2}}

Let us describe the unique $G$-equivariant connection on the
vector bundles associated with the principal $H$-bundle
\beq
\sut~\xr{\Sp(\utwo\times\uo)}~\C P^2 \ .
\label{Stiefelsut}\eeq
The projective plane can be covered by three patches, and on one of
these patches we choose complex coordinates
\beq
Y~:=~\begin{pmatrix} y^1\\[4pt] y^2\end{pmatrix} \qquad
\mbox{and} \qquad Y^\dag\=\big(\yb^1~\yb^2\big)
\label{YU0}\eeq
with $Y^\dag\,Y=\yb^i\,y^i$ and $i=1,2$. Introduce the column one-form
\beq
\betab~:=~\begin{pmatrix}\betab^1\\[4pt]\betab^2\end{pmatrix} \qquad
\mbox{with} \qquad \betab^i\=\frac1{\gamma}~\diff\yb^i-\frac{\yb^i}
{\gamma^2\,(\gamma+1)}\,y^j~\diff\yb^j \ ,
\label{betacomps}\eeq
where
\beq
\gamma:=\sqrt{1+\yb^i\,y^i} \ .
\label{gammadef}\eeq
The $(1,0)$-forms $\beta^i$ and the $(0,1)$-forms $\betab^i$
constitute a $G$-equivariant basis for the complex vector spaces of
forms of type $(1,0)$ and $(0,1)$ on $\CP^2$, respectively, and give
the horizontal components of a flat connection $A_0$ tangent to the
base of the bundle (\ref{Stiefelsut})~\cite{LPS3}.

Consider the $G$-equivariant field given by 
\beq
a=-\frac1{2\gamma^2}\,\big(y^i~\diff\yb^i-\yb^i~\diff y^i\big) \
.
\label{CP2monfield}\eeq
The one-form (\ref{CP2monfield}) is the $\uoL$-valued monopole
potential on $\C P^2$ which can be described as the canonical abelian
connection on the Hopf bundle
\beq
S^5=\ut\,\big/\,\utwo~\xr{\uo}~\C P^2 \ .
\label{S5Hopfbundle}\eeq
The complex line bundle $\Lcal\to\C P^2$ associated with the principal
$\uo$-bundle (\ref{S5Hopfbundle}) is the monopole bundle over $\C
P^2$ which we take to be endowed with the same $\uoL$-connection
$a$. It is a representative of the isomorphism class in $\HQ^1(\C
P^2;\uo)\cong\Z$ corresponding to the abelian field strength
\beq
f_{\uoL}~:=~\diff a\=
\betab^1\wedge\beta^1+\betab^2\wedge\beta^2 \ .
\label{CP2moncurvexpl}\eeq
Higher degree monopole bundles $\Lcal_{m/2}:=\big(\Lcal^{\otimes
  m}\big)^{1/2}$ are endowed with the connection $\frac m2\,a$. These
bundles  are associated to higher weight irreducible representations
$\underline{(m)}$ of the fibres of (\ref{S5Hopfbundle}) but only exist
globally when $m$ is even, as only then is the first Chern number
${\frac m2}$ an integer. Nevertheless, odd values of $m$ are necessary
for construction of the $\urm(2)$ bundle $\Qcal$ with curvature
$F_{\utwoL}$ below and, as we shall see, for considering invariant
spinors. The justification for calling $\frac m2$ the ``monopole
charge'' is explained in \S{\ref{ChernCharacter}}.
The monopole field strength of charge $\frac
m2$ is a $(1,1)$-form proportional to the canonical K\"ahler two-form
on $\C P^2$ defined by
\beq
\omega=\im R^2\,\big(\beta^1\wedge\betab^1+
\beta^2\wedge\betab^2\big) \ ,
\label{KahlerCP2}\eeq
where $R$ is the radius of the linearly embedded projective line $\C
P^1\subset\C P^2$ whose homology class is Poincar\'e dual to the
cohomology class of (\ref{CP2moncurvexpl}).

Consider now the $G$-equivariant field $B\in\utwoL$ defined by
\beq
B=\frac1{\gamma^2}\,\big(-\mbox{$\frac12$}~\diff(Y^\dag\,Y)~
\Idd_2+\Yb~
\diff\Yb^\dag+\Lambda~\diff\Lambda\big) \ , 
\label{U2instfield}\eeq
where
\beq
\Lambda:=\gamma~\Idd_2-\frac1{\gamma+1}~Y\,Y^\dag \ .
\label{Lambdadef}\eeq
The one-forms $B-\frac12\,\tr(B)~\Idd_2$ and $a$ on $\C P^2$ give the
vertical components of $A_0$ with values in the tangent space ${\rm
  su}(2)\oplus\uoL$ to the fibre of the bundle (\ref{Stiefelsut}), and
together with the forms (\ref{betacomps}) they obey the Cartan-Maurer
equations
\beq
\diff\betab+B\wedge\betab+2a\wedge\betab\=0 \qquad \mbox{and}
\qquad \diff\beta-B\wedge\beta-2a\wedge\beta\=0 \ .
\label{diffbetaBa0}\eeq
The $\utwoL$-valued curvature
\beq
F_{\utwoL}~:=~\diff B+B\wedge B\=\betab\wedge\beta^\top\=
F_{\suL}+\mbox{$\frac12$}\,f_{\uoL}~\Idd_2
\label{Fu2reduce}\eeq
can be expressed in terms of the abelian field strength
(\ref{CP2moncurvexpl}) and the curvature
\beq
F_{\suL}~:=~\begin{pmatrix}\mbox{$\frac12$}\,\big(
\betab^1\wedge\beta^1-\betab^2\wedge\beta^2\big)&
\betab^1\wedge\beta^2\\[4pt]\betab^2\wedge\beta^1&
-\mbox{$\frac12$}\,\big(
\betab^1\wedge\beta^1-\betab^2\wedge\beta^2\big)\end{pmatrix}\=
\diff B_{(1)}+B_{(1)}\wedge B_{(1)}
\label{Fsu2inst}\eeq
of the gauge potential $B_{(1)}=B-\frac12\,a~\Idd_2\in\suL$. The
one-form $B_{(1)}$ is the $\suL$-valued one-instanton field on $\C
P^2$ constructed from the canonical connection on a principal ${\rm
  U}(2)$-bundle as described in \S{\ref{ChernCharacter}}. Its curvature $F_{\suL}$
is, at least locally, a $(1,1)$-form on $\C P^2$. Higher rank instanton bundles
$\Ical_n$ are endowed with $G$-equivariant one-instanton connections
$B_{(n)}\in{\rm su}(n+1)$ and fibre spaces in $(n+1)$-dimensional
irreducible representations of the fibres of this bundle. As explained in \S{\ref{ChernCharacter}}, the
bundle $\Ical_n$ is only globally defined for even values of $n$.
For a given representation $\underline{(n,m)}$ of
the holonomy group $H=\Sp(\utwo\times\uo)$, the corresponding
homogeneous vector bundle (\ref{indhermbungen}) is given by
(\ref{Hcalnmdef}) and can be identified with
$\Ical_n\otimes\Lcal_{m/2}$.

\subsection{Invariant gauge fields\label{invgaugefields}}

To determine the generic form of a $G$-equivariant connection one-form
$\ca$ on the vector bundle $\Ecal^{k,l}\to \man$, let us assume for
simplicity that $M$ (and hence $\man$) is a complex manifold. We
decompose the space $\Omega^{0,1}(\End(\Ecal^{k,l}))^G$ using the
Whitney sum (\ref{Ecalklisotopical}). By Schur's lemma,
corresponding to each weight $(n,m)\in\quiver_{k,l}$ there is a
``diagonal'' subspace
\beq
\big(\Omega^{0,1}(\End(E_{n,m}))\otimes\Idd_{n+1}\big)~\oplus~
\big( \Idd_{p_{n,m}}\otimes\Omega^{0,1}(\End(\Hcal_{n,m}))^G\,\big) \
,
\label{diagsubsp}\eeq
in which we can choose a connection $A^{n,m}$ on the bundle
$E_{n,m}\to M$ twisted by a $G$-equivariant connection on the
homogeneous vector bundle $\Hcal_{n,m}\to\C P^2$ constructed from the
gauge potentials $a$ and $B_{(1)}$ of \S\ref{canconnCP2}. To each
weight morphism (\ref{CP2arrows}) there is an ``off-diagonal''
subspace
\beq
\Omega^0\big(\Hom(E_{n,m},E_{{n\pm1,m+3}})\big)~\otimes~
\Omega^{0,1}\big(\Hom(\Hcal_{n,m},\Hcal_{n\pm1,m+3})\big)^G \ ,
\label{offdiagsubsp}\eeq
in which we twist the Higgs fields $\phi_{n,m}^\pm$ by suitable
invariant $(n\pm1+1)\times(n+1)$ matrix-valued $(0,1)$-forms built
from the basis $(0,1)$-forms $\betab^i$ spanning $\Omega^{0,1}(\C
P^2)^G$ that were constructed in \S\ref{canconnCP2}. Thus the
condition of $G$-equivariance uniquely dictates the form of the gauge
connection $\ca$ in $(n+1)\,p_{n,m}\times(n\pm1+1)\,p_{n\pm1,m+3}$
blocks.

To appropriately assemble the invariant $(0,1)$-forms into rectangular
block matrices, we will use the Biedenharn basis for the irreducible
representations $\underline{C}^{k,l}$ of $\sut$. The complete set
of $d^{k,l}$ orthonormal vectors in this basis set are denoted
$\big|\noverq,m\big\rangle$, and are labelled by the isospin quantum
numbers $n=2I$, $q=2I_z$ and the hypercharge $m=3Y$. These states
define the spin $\frac n2$ representation of the isospin subgroup
$\su\subset\sut$ and are hypercharge eigenstates with the matrix
elements
\bea
H_{\alpha_1}\big|\noverq\,,\,m\big\rangle&=&
q\,\big|\noverq\,,\,m\big\rangle \ , \label{Izrep}\\[4pt]
E_{\pm\,\alpha_1}\big|\noverq\,,\,m\big\rangle&=&
\mbox{$\frac12$}\,\sqrt{(n\mp q)\,(n\pm q+2)}~\big|\noverqpmt\,,\,m
\big\rangle \ , \label{Ipmrep}\\[4pt]
H_{\alpha_2}\big|\noverq\,,\,m\big\rangle&=&m\,\big|\noverq\,,\,m
\big\rangle \ , \label{Yrep}\\[4pt]
E_{\alpha_2}\big|\noverq\,,\,m\big\rangle&=&
E^+_{\alpha_2}\big|\noverq\,,\,m\big\rangle+
E^-_{\alpha_2}\big|\noverq\,,\,m\big\rangle \label{Fprep}\\[4pt] &:=&
\sqrt{\mbox{$\frac{n-q+2}{2(n+1)}$}}~\Lambda_{k,l}^+(n,m)\,
\big|\npoverqmo\,,\,m+3\big\rangle+
\sqrt{\mbox{$\frac{n+q}{2(n+1)}$}}~\Lambda_{k,l}^-(n,m)\,
\big|\nmoverqmo\,,\,m+3\big\rangle \ , \nonumber\\[4pt]
E_{\alpha_1+\alpha_2}\big|\noverq\,,\,m\big\rangle&=&
E^+_{\alpha_1+\alpha_2}\big|\noverq\,,\,m\big\rangle+
E^-_{\alpha_1+\alpha_2}\big|\noverq\,,\,m\big\rangle \label{Fmrep}\\[4pt]
&:=&
\sqrt{\mbox{$\frac{n+q+2}{2(n+1)}$}}~\Lambda_{k,l}^+(n,m)\,
\big|\npoverqpo\,,\,m+3\big\rangle+
\sqrt{\mbox{$\frac{n-q}{2(n+1)}$}}~\Lambda_{k,l}^-(n,m)\,
\big|\nmoverqpo\,,\,m+3\big\rangle \ , \nonumber
\eea
where
\bea
\Lambda_{k,l}^+(n,m)&=&\mbox{$\frac1{\sqrt{n+2}}~
\sqrt{\big(\frac{k+2l}3+\frac n2+\frac
    m6+2\big)\,\big(\frac{k-l}3+\frac n2+\frac m6+1\big)\,\big(
\frac{2k+l}3-\frac n2-\frac m6\big)}$} \ , \nonumber\\[4pt]
\Lambda_{k,l}^-(n,m)&=&\mbox{$\frac1{\sqrt n}~
\sqrt{\big(\frac{k+2l}3-\frac n2+\frac
    m6+1\big)\,\big(\frac{l-k}3+\frac n2-\frac m6\big)\,\big(
\frac{2k+l}3+\frac n2-\frac m6+1\big)}$} \ .
\label{lambdaklnm}\eea
The latter constants are defined for $n>0$ and we set
$\Lambda_{k,l}^-(0,m):=0$. The analogous relations for $E_{-\alpha_2}$
and $E_{-\alpha_1-\alpha_2}$ can be derived by hermitean conjugation
of (\ref{Fprep}) and (\ref{Fmrep}), respectively.

For a fixed weight $(n,m)\in\quiver_{k,l}$, we write the
one-instanton connection $B_{(n)}=B_{n,m}$ in the $(n+1)$-dimensional
irreducible representation of $\su$ as
\bea
B_{n,m}&:=&B^{11}\,H_{\alpha_1}+B^{12}\,E_{\alpha_1}-\big(B^{12}\,
E_{\alpha_1}\big)^\dag \nonumber\\[4pt]
&=& \sum_{q\in\rel_n}\,\Big(q\,B^{11}\,\big|
\noverq\,,\,m\big\rangle\big\langle\noverq\,,\,m\big|+
\mbox{$\frac12$}\,B^{12}\,\sqrt{(n-q)\,(n+q+2)}~\big|
\noverqpt\,,\,m\big\rangle\big\langle\noverq\,,\,m\big|\nonumber\\
&& \hspace{2cm} -\,\mbox{$\frac12$}\,
\overline{B^{12}}\,\sqrt{(n+q)\,(n-q+2)}~\big|
\noverqmt\,,\,m\big\rangle\big\langle\noverq\,,\,m\big|\Big)
\label{Bndef}\eea
where $\rel_n:=\{-n,-n+2,\dots,n-2,n\}$, and $B^{ij}$ are the matrix
elements of the $\suL$-valued instanton connection
$B_{(1)}=B-\frac12\,a~\Idd_2$. The monopole potential is represented
in this basis by $\frac12\,a\,H_{\alpha_2}$. Denote by
\beq
\Pi_{n,m}:=\sum_{q\in\rel_n}\,\big|
\noverq\,,\,m\big\rangle\big\langle\noverq\,,\,m\big|
\label{Pinmdef}\eeq
the projection of $\underline{C}^{k,l}\big|_H$ onto the irreducible
representation $\underline{(n,m)}$ of $H=\su\times\uo$. We further
write
\beq
\betab^{\pm}\=
\betab^1\,E^\pm_{\alpha_1+\alpha_2}+\betab^2\,E^\pm_{\alpha_2} 
\=\sum_{(n,m)\in\quiver_{k,l}}\, \betab^{\pm}_{n,m} \ ,
\label{betapmdef}
\eeq
where
\bea
\betab^{\pm}_{n,m}&:=&
\frac{\Lambda_{k,l}^\pm(n,m)}{\sqrt{2(n+1)}}~
\sum_{q\in\rel_n}\,\Big(\sqrt{n\pm q+1\pm1}~\betab^1~
\big|\npmoverqpo\,,\,m+3\big\rangle\big\langle\noverq\,,\,
m\big| \nonumber\\ && \hspace{4cm}
+\,\sqrt{n\mp q+1\pm1}~\betab^2~
\big|\npmoverqmo\,,\,m+3\big\rangle\big\langle\noverq\,,\,
m\big|\Big)
\label{betanmdef}\eea
are the $(n\pm1+1)\times(n+1)$ matrix blocks of $G$-equivariant
elementary bundle morphisms between $\Hcal_{n,m}$ and
$\Hcal_{n\pm1,m+3}$, together with their hermitean conjugates
$\beta_{n,m}^\pm:={{\betab}^\pm_{n,m}}{}^\dag$.

By introducing the projection $\pi_{n,m}$ onto the sub-bundle
$E_{n,m}\to M$, the anti-hermitean $G$-equivariant gauge
connection $\ca$ on the bundle (\ref{Ecalklisotopical}) over
$M\times\C P^2$ can be written as
\bea
\ca&=&\sum_{(n,m)\in\quiver_{k,l}}\,\Big(A^{n,m}\otimes
\Pi_{n,m}+\pi_{n,m}\otimes\big(B_{n,m}+\mbox{$\frac m2$}\,
a  ~\Pi_{n,m}\big)
\label{calACP2Cklrep}\\ && \hspace{3cm}
+\,\phi_{n,m}^+\otimes\betab^+_{n,m}+\phi_{n,m}^-\otimes
\betab_{n,m}^--
{\phi_{n,m}^+}^\dag\otimes\beta^+_{n,m}-{\phi_{n,m}^-}^\dag
\otimes\beta_{n,m}^-\Big) \ . \nonumber
\eea
Note that when $j_+=\frac k2$, one has $\Lambda^+_{k,l}(n,m)=0$ for
all $j_-$ and when $j_-=0$, one has $\Lambda^-_{k,l}(n,m)=0$ for all
$j_+$, so the corresponding fields $\betab^\pm_{n,m}$ and
$\phi^\pm_{n,m}$ also vanish. These two cases correspond respectively
to the rightmost and bottom edges in (\ref{Figquiver}). We can thus
associate the Higgs fields $\phi^+_{n,m}$ with the horizontal links in
(\ref{Figquiver}) and $\phi^-_{n,m}$ with the vertical links. Then
there are a total of $2k\,l+k+l$ independent fields $\phi^\pm_{n,m}$.

The matrix elements of the curvature two-form
\beq
\cf=\diff\ca+\ca\wedge\ca
\label{cfca}\eeq
are straightforwardly computed in the Biedenharn basis by using
(\ref{diffbetaBa0})--(\ref{Fsu2inst})~\cite{LPS3}. For each
weight $(n,m)\in\quiver_{k,l}$ one finds the diagonal matrix elements
\bea
\cf^{n,m\,;\,n,m}&=&F^{n,m}\otimes\Idd_{n+1}+
\big(\Idd_{p_{n,m}}-{\phi_{n,m}^+}^\dag\,
\phi_{n,m}^+\big)\otimes\big(\beta_{n,m}^+\wedge\betab_{n,m}^+\big)
\nonumber\\ && +\,\big(
\Idd_{p_{n,m}}-{\phi_{n,m}^-}^\dag\,\phi_{n,m}^-\big)\otimes\big(
\beta_{n,m}^-\wedge\betab_{n,m}^-\big) \nonumber\\ &&
+\,\big(\Idd_{p_{n,m}}-\phi_{n-1,m-3}^+\,{\phi_{n-1,m-3}^+}^\dag
\,\big)\otimes\big(\betab_{n-1,m-3}^+\wedge\beta_{n-1,m-3}^+\big)
\nonumber\\ &&
+\,\big(\Idd_{p_{n,m}}-\phi_{n+1,m-3}^-\,{\phi_{n+1,m-3}^-}^\dag
\,\big)\otimes\big(\betab_{n+1,m-3}^-\wedge\beta_{n+1,m-3}^-\big)
\label{curvdiagCklrep}\eea
where $F^{n,m}=\diff A^{n,m}+A^{n,m}\wedge A^{n,m}$ is the curvature
of the vector bundle $E_{n,m}\to M$, while the non-vanishing
off-diagonal matrix elements are given by
\beq
\cf^{n\pm1,m+3\,;\,n,m}\=D\phi_{n,m}^\pm\wedge
\betab_{n,m}^\pm~:=~\big(\diff\phi_{n,m}^\pm+A^{n\pm1,m+3}\,
\phi_{n,m}^\pm-\phi_{n,m}^\pm\,A^{n,m}\big)\wedge\betab_{n,m}^\pm
\label{curvoffdiagCklrep}\eeq
and
\bea
\cf^{n+1,m+3\,;\,n+1,m-3}&=&\big(\phi_{n,m}^+\,\phi_{n+1,m-3}^--
\phi_{n+2,m}^-\,\phi_{n+1,m-3}^+\big)\otimes\big(\betab_{n,m}^+\wedge
\betab^-_{n+1,m-3}\big) \ , \label{curvholrelsCklrep}\\[4pt]
\cf^{n+1,m+3\,;\,n-1,m+3}&=&\big(\phi_{n,m}^+\,{\phi_{n,m}^-}^\dag-
{\phi_{n+1,m+3}^-}^\dag\,\phi_{n-1,m+3}^+\big)\otimes
\big(\betab^+_{n,m}\wedge\beta^-_{n,m}\big)
\label{curvnonholrelsCklrep}\eea
along with their hermitean conjugates
$\cf^{r,s\,;\,n,m}=-(\cf^{n,m\,;\,r,s})^\dag$ for
$(r,s)\neq(n,m)$. The matrix elements
(\ref{curvoffdiagCklrep}) define bi-fundamental covariant derivatives
$D\phi_{n,m}^\pm$ of the Higgs fields. The matrix one-form products
appearing above are written out explicitly in
\S{\ref{betamatrixprods}}.

\subsection{Invariant spinor fields\label{InvSpinors}}

Let $M$ be a complex manifold, so that $d=\dim_\R(M)$ is even, and let
$K=\bigwedge_\C^{d/2}(T^*M)$ be its canonical line bundle. If
$c_1(K)=0$ mod~$2$ then $M$ is a spin manifold, while generically
$K$ determines a canonical spin$^c$-structure on $M$. The
corresponding spin$^c$-bundles are denoted
$\underline{\Delta}\,(M)\cong \bigwedge^{0,\bullet}(TM)$
and are obtained by twisting the usual spinor bundles associated to
the principal $\spin(d)$-bundle $P_{\spin(d)}\to M$ by $K^{-1/2}$. By
spinor fields on $M$ or $\man$ we shall always refer to sections of
such spin$^c$-bundles.

The equivariant dimensional reduction of massless Dirac spinors on
$M\times\C P^2$ is defined with respect to (twisted) symmetric
fermions on $M$. They act as intertwining operators connecting induced
representations of the holonomy group $H=\su\times\uo$ in the
$\urm(p)$ gauge group, and also in the twisted spinor module
$\underline{\Delta}\,(M)$ which admits the isotopical decomposition
\beq
\underline{\Delta}\,(M)\=\bigoplus_{(n,m)\in\quiver_{k,l}}\,
\Delta_{n,m}\otimes\,
\underline{(n,m)} \qquad\mbox{with}\quad \Delta_{n,m}\={\rm
  Hom}^{~}_{H}\big(\,\underline{(n,m)}\,,\,
\underline{\Delta}\,(M)\big)
\label{spinmoddecomp}\eeq
obtained by restricting $\underline{\Delta}\,(M)$ to representations
of $H\subset\spin^c(d)=\spin(d)\times_{\Z_2}\,\uo$. By Frobenius
reciprocity, the multiplicity spaces may be identified as
\beq
\Delta_{n,m}=\bigl(\,\underline{\Delta}\,(M)^\vee\otimes
\Omega^0(\Hcal_{n,m})\bigr)^G \ ,
\label{multspHomG}\eeq
and hence the isotopical decomposition (\ref{spinmoddecomp}) is
realized explicitly by constructing symmetric fermions
on $M$ as $\sut$-invariant spinors on $M \times\C P^2$. They are
associated with the eigenspinors of the twisted Dirac operator on $\C
P^2$, which we describe in some detail.

There is a global obstruction to defining spinors on $\CP^2$, but a
spin$^c$ structure can be defined by twisting the usual spinor bundle
with half-integer powers of the monopole line bundle $\Lcal$. At the
level of the twisted Dirac operator, this can be achieved by changing
the coupling to the $\uo$ component of the invariant gauge potential
(\ref{calACP2Cklrep}), and we therefore propose this as a method for
describing spinors globally on $\CP^2$. The complete spectrum of the
Dirac operator on $\CP^2$, coupled to arbitrary instanton and monopole
backgrounds, was worked out in~\cite{BDHom}. The eigenspinors for an
arbitrary monopole background, without instantons, were constructed in
the context of the fuzzy projective plane $\CP_F^2$
in~\cite{DiraconFuzzyCP2},\footnote{Monopole line bundles on $\CP_F^2$
are also discussed in~\cite{CWSW}.} while the number of zero modes in a rank
two instanton background with arbitrary monopole charge was originally
computed in~\cite{BDCN}. The number of spinor harmonics in a generic
instanton background and with arbitrary monopole number is computed in
\S{\ref{ASItheorem}}. In this section we will restrict
attention to zero modes of the Dirac operator on $\CP^2$.

Recall that the pairs $(n,m)\in\quiver_{k,l}$ appearing in
(\ref{calACP2Cklrep}) have the same even/odd integer parity. Suppose
we try to write down a Dirac operator acting on spinors on $M\times
\CP^2$ coupled to the gauge connection (\ref{calACP2Cklrep}),
transforming under some fixed representation $\rho$ of the subgroup
$\prod_{(n,m)\in\quiver_{k,l}}\,\SU(p_{n,m})$ of the gauge group and
under the same weights of $\urm(2)$ as those occurring in the
decomposition (\ref{calACP2Cklrep}). Such spinors couple to
topologically non-trivial $\SU(2)\times\urm(1)$ gauge potentials on
$\CP^2$. Then there will be an inconsistency because the index of the
Dirac operator is fractional, reflecting the fact
that spinor fields are never globally well-defined on $\CP^2$ in these
backgrounds. For a generic $\SU(2)\times\urm(1)$ representation
$\underline{(n,m)}$, the index is calculated in \S{\ref{ASItheorem}}
to be
\beq
\nu_{b;n}=\mbox{$\frac12$}\,(n+1)\,(b+1)\,(n+b+2)
\label{nunb}
\eeq
where $b=\frac {m-n-3}2$. If $n$ and $m$ have the same parity then
$b$ is not an integer.

To avoid this obstruction, we modify the Dirac operator by twisting it
with a half-integer power $\Lcal_{\ct}$, $\ct\in\Z+\frac 12$ of the
monopole line bundle $\Lcal$. The Dirac operator acting on
four-component twisted spinor fields
$\chi_{n,m;\ct}\in\Omega^{0,\bullet}(\Hcal_{n,m+2\,\ct})$ on $\CP^2$,
coupled to the rank~$n+1$ instanton connection and magnetic monopole
potential of charge $\frac m2 +\ct$, is then
\beq
\Dirac_{\CP^2}=\sum_{(n,m)\in\quiver_{k,l}}\,\Big( \pa_{\CP^2}\,
\Pi_{n,m}+ B_{n,m}\kern -22pt/ \kern+16pt + \bigl(\mbox{$\frac m2
  +\ct$} \bigr)\; a\kern -6pt/\; \Pi_{n,m}\Big) \ ,
\label{DiracCP2}\eeq
where $\pa_{\CP^2}$ is the naive Dirac operator on $\CP^2$ involving
only the spin connection. The index for weight $(n,m)$ is again given
by (\ref{nunb}), but now this is an integer when $b=\ct +\frac
{m-n-3}2$. Since $b$ depends only on $(n,m)$, and on $\ct$ which is
half-integer, we denote it by $b_{n,m+c}$ where $c=2\,\ct$ is an
odd integer. Then the index for a given irreducible $\urm(2)$-module
$\underline{(n,m)}$ is
\beq
\nu_{n,m}=\mbox{$\frac12$}\,(n+1)\,(b_{n,m+c}+1)\,(n+b_{n,m+c}+2) \ .
\label{nunm}
\eeq

For fixed $\ct$ we shall denote the positive/negative chirality
zero modes of the Dirac operator (\ref{DiracCP2}) by
$\chi^\pm_{n,m}\in\C^2$. From the explicit construction in
\cite{DiraconFuzzyCP2}, it is known that for $n=0$ the index coincides
with the total number of zero modes, so either all spinor harmonics have
positive chirality or all have negative chirality. We will assume that
the same property is true for all $n\ge1$. Although we do not have a
rigorous proof, this seems plausible given the natural identification
of the virtual zero mode eigenspaces of $\Dirac_{\CP^2}$ with
irreducible representations of $\SU(3)$ discussed in
\S\ref{ASItheorem}.\footnote{In any case, if this is not true then the
  same qualitative conclusions below will hold, but the notation would
  have to be modified to incorporate the extra spinor harmonics.} With
this assumption, in a suitable basis there are chiral decompositions
\beq
\Dirac_{\CP^2}=
\sum_{(n,m)\in\quiver_{k,l}}\,\begin{pmatrix}
0 & \Dirac_{n,m}^+ \\ \Dirac_{n,m}^- & 0 \end{pmatrix}
\label{Diracchiraldecomp}\eeq
of (\ref{DiracCP2}) into twisted Dolbeault-Dirac operators
$\Dirac_{n,m}^\pm$, such that the index (\ref{nunm}) is the virtual
dimension of the vector space
$\ker\big(\Dirac_{n,m}^+\big)\ominus\ker\big(\Dirac_{n,m}^-\big)$. Then
$\chi_{n,m}^\pm\neq0$ only when $(n,m)\in\quiver_{k,l}^\pm$, where
\beq
\quiver_{k,l}^\pm:=\big
\{(n,m)\in\quiver_{k,l}~\big|~\pm\,\nu_{n,m}>0\big\} \ .
\label{weightklpm}\eeq
We fix a basis of chiral/antichiral spinor harmonics
$\chi_{n,m;\ell}^\pm\in\ker\big(\Dirac_{n,m}^\pm\big)$,
$\ell=1,\dots,|\nu_{n,m}|$ for each weight
$(n,m)\in\quiver_{k,l}^\pm$. They transform for each $\ell$ in the
$(n+1)$-dimensional irreducible representation of the isospin subgroup
$\su\subset H$ of the holonomy group.

We can now use (\ref{multspHomG}) to take tensor products of the Dirac
zero modes on $\CP^2$ with (twisted) Dirac spinors
$\psi_{n,m;\ell},\widetilde\psi_{n,m;\ell}\in
\Omega^{0,\bullet}\big(\rho(E_{n,m})\big)$, $\ell=1,\dots,|\nu_{n,m}|$
on $M$ to produce fermion fields
\bea
\Psi_{n,m}^+&=&\sum_{\ell=1}^{\nu_{n,m}}\,\psi_{n,m;\ell}
\otimes \chi^+_{n,m;\ell} \qquad \mbox{and} \qquad \Psi_{n,m}^-\=0
\qquad \mbox{for} \quad (n,m)\in\quiver_{k,l}^+ \ , \nonumber\\[4pt]
\Psi_{n,m}^-&=&\sum_{\ell=1}^{|\nu_{n,m}|}\,
\widetilde\psi_{n,m;\ell}\otimes \chi^-_{n,m;\ell} \qquad \mbox{and}
\qquad \Psi_{n,m}^+\=0 \qquad \mbox{for} \quad (n,m)\in\quiver_{k,l}^-
\ .
\label{Psipsichi}\eea
Note that the spinors $\Psi_{n,m}^\pm$ are not chiral on
$M\times\CP^2$. From these fields we construct a $G$-equivariant Dirac
spinor field on $\man=M\times \CP^2$ as
\beq
\Psi\=\left(\begin{matrix} \Psi^+ \\ \Psi^-  \end{matrix} \right)\=
\bigoplus_{(n,m)\in\quiver_{k,l}}\,\left(\begin{matrix}
\Psi^+_{n,m} \\ \Psi^-_{n,m}\end{matrix}\right) \ .
\label{Psiequiv}\eeq

\bigskip

\section{Quiver gauge theory\label{Eqgaugeth}}

\noindent
In this section we shall work out the equivariant dimensional
reduction of pure massless Yang-Mills-Dirac theory on the manifold
(\ref{M2ntimesSU3H}). We will emphasise the roles played by the
$\su$-instanton and $\uo$-monopole background fields on $\C P^2$,
particularly how they affect the vacuum structure of the quiver gauge
theory corresponding to (\ref{Figquiver}). We shall also compare the
induced equivariant gauge theory on $M$ with that obtained via
dimensional reduction over $\C P^1$~\cite{Dolan09}.

\subsection{Reduction of the Yang-Mills action}

We endow the manifold $M$ with local real coordinates $x=(x^\m
)\in\R^{d}$, where the indices $\m ,\n,\ldots$ run through $1,\ldots
,d$. The metric
\beq
\diff s^2=\Gcal_{AB}~\diff x^{A}\otimes\diff x^{B}
\label{metrichat}\eeq
on $\man=M\times \C P^2$ will be taken to be the direct product of a
chosen riemannian metric on $M$ and the canonical ${\rm
  SU}(3)$-symmetric K\"ahler metric on $\C P^2$ corresponding to the
two-form (\ref{KahlerCP2}), where the indices $A,B,\dots$ run over
$1,\dots,d+4$. Working in the basis $\beta^i$, $\betab^i$ of
invariant forms on $\C P^2$ and in the coordinates above, it takes the
form
\beq\label{metric3}
\diff s^2 =G_{\mu\nu}\ \diff x^\mu\otimes\diff x^\nu +
2R^2\,\big(\beta^1\otimes\betab^1+\beta^2\otimes\betab^2\big) \ .
\eeq
The line element (\ref{metric3}) has mass dimension~$-2$.

The pure Yang-Mills lagrangian on $\man=M\times\C P^2$ is given by
\bea
L^{k,l}_{\rm YM}&=&-\frac{1}{4\tilde g^2}\,\sqrt{|\Gcal|}~
\tr^{~}_{p\times p}\ \cf_{AB}\,\cf^{AB} \nonumber\\[4pt]
&=&-\frac{1}{4\tilde g^2}\,\sqrt{|\Gcal|}~
\tr^{~}_{p\times p} \bigg[\,\cf_{\mu\nu}\,\cf^{\mu\nu}+ 
\frac{1}{2R^2}\,G^{\mu\nu}\,
\big(\cf_{\mu i}\,\cf_{\nu \bar \imath}+\cf_{\mu\bar\imath}\,
\cf_{\nu i}\big) \nonumber\\ && \hspace{4cm}
-\,\frac1{2R^4}\,\big(|\cf_{1\bar1}|^2+
|\cf_{2\bar2}|^2+2|\cf_{1\bar2}|^2+2|\cf_{12}|^2\big)\,\bigg]
\label{lagrprod}\eea
where we use the matrix notation $|\cf|^2:=
\frac12\,(\cf^\dag\,\cf+\cf\,\cf^\dag)$, and $i=1,2$ labels
components along $\C P^2$ in the basis used in (\ref{metric3}). The
$(d+4)$-dimensional $\uk$ Yang-Mills coupling constant $\tilde g$ has
the standard mass dimension $-\frac d2$ in order to make
(\ref{lagrprod}) dimensionless. We substitute
(\ref{curvdiagCklrep})--(\ref{curvnonholrelsCklrep}) into
(\ref{lagrprod}), and take the trace over the representation space
$\underline{(n,m)}$ for each weight $(n,m)\in\quiver_{k,l}$ making use
of the identities of~\S\ref{betamatrixprods}. We then integrate over
$\C P^2$ using the normalization $\int_{\CP^2}\, \beta_{\rm vol}=1$,
where $\beta_{\rm vol}:=
\frac1{2\pi^2}\,\beta^1\wedge\betab^1\wedge\beta^2\wedge\betab^2$ is
the unit volume form of $\CP^2$.

After some calculation and rescaling $\phi_{n,m}^\pm\to
\Lambda^\pm_{k,l}(n,m)^{-1}\,\phi_{n,m}^\pm$, one finds that the
dimensional reduction of the corresponding Yang-Mills action
\beq
S^{k,l}_{\rm YM}:=\int_{M\times {\C P^2}}\,
{\diff^{d+4}}x~L_{\rm YM}^{k,l}
\label{SklYMdef}\eeq
is given by
\bea
S^{k,l}_{\rm YM}&=&
\frac{\big(\pi\,R^2\big)^2}{2\tilde g^2}\, 
\int_{M}\,\diff^{d}x~\sqrt{|G|}\ \sum_{(n,m)\in\quiver_{k,l}}\,
\tr_{p_{n,m}\times p_{n,m}}\bigg[\,(n+1)\,\big(F^{n,m}
\big)^\dag_{\mu\nu}\,\big(F^{n,m}\big)^{\mu\nu} \nonumber  \\ &&
+\, \frac{n+2}{2R^2}\,\big(D_\mu\phi_{n,m}^+
\big)^\dag\,\big(D^\mu\phi_{n,m}^+\big) +\frac{n+1}{2R^2}\,
\big(D_\mu\phi_{n-1,m-3}^+\big)\,\big(D^\mu\phi_{n-1,m-3}^+
\big)^\dag \nonumber \\ && +\,\frac{n}{2R^2}\,\big(D_\mu\phi_{n,m}^-
\big)^\dag\,\big(D^\mu\phi_{n,m}^-\big)+\frac{n+1}{2R^2}\,
\big(D_\mu\phi_{n+1,m-3}^-\big)\,\big(D^\mu\phi_{n+1,m-3}^-
\big)^\dag \nonumber \\ && +\,\frac{n+2}{2R^4}\,\Big(
\Lambda_{k,l}^+(n,m)^2~\Idd_{p_{n,m}}-\phi_{n,m}^+{}^\dag\,
\phi_{n,m}^+\Big)^2+\frac{n}{2R^4}\,\Big(
\Lambda_{k,l}^-(n,m)^2~\Idd_{p_{n,m}}-\phi_{n,m}^-{}^\dag\,
\phi_{n,m}^-\Big)^2 \nonumber \\ && +\,\frac{(n+1)^2}{2n\,R^4}\,
\Big(\Lambda_{k,l}^+(n-1,m-3)^2~\Idd_{p_{n,m}}-\phi_{n-1,m-3}^+{}\,
\phi_{n-1,m-3}^+{}^\dag\Big)^2 \label{SklYM} \\ && +\,
\frac{(n+1)^2}{2(n+2)\,R^4}\,
\Big(\Lambda_{k,l}^-(n+1,m-3)^2~\Idd_{p_{n,m}}-\phi_{n+1,m-3}^-{}\,
\phi_{n+1,m-3}^-{}^\dag\Big)^2 \nonumber \\ && +\,
\frac {n\,(n+2)}{2(n+1)\,R^4}\,\Big|\phi_{n,m}^+\,\phi_{n,m}^-{}^\dag-\frac{
\Lambda_{k,l}^+(n,m)\,\Lambda_{k,l}^-(n,m)}
{\Lambda_{k,l}^+(n-1,m+3)\,\Lambda_{k,l}^-(n+1,m+3)}\,
\phi_{n+1,m+3}^-{}^\dag\,\phi_{n-1,m+3}^+\Big|^2 \nonumber \\ && +\,
\frac{n+3}{6R^4}\,\Big|\phi_{n,m}^+\,\phi_{n+1,m-3}^--
\frac{\Lambda_{k,l}^+(n,m)\,\Lambda_{k,l}^-(n+1,m-3)}
{\Lambda_{k,l}^+(n+1,m-3)\,\Lambda_{k,l}^-(n+2,m)}\,
\phi_{n+2,m}^-\,\phi_{n+1,m-3}^+\Big|^2\,\bigg] \ . \nonumber
\eea
From (\ref{curvoffdiagCklrep}) it follows that the $\uo$ factor in the
structure group $\uk\cong\uo\times\SU(p)$ does not enter the
bicovariant derivatives of the rectangular scalar fields
$\phi_{n,m}^\pm$. We can therefore restrict to gauge group $\SU(p)$,
and the decomposition (\ref{gaugegroupbreak}) is then modified to
\beq
\SU(p)~\longrightarrow~\uo^{(k+1)\,(l+1)-1}\times
\prod_{(n,m)\in\quiver_{k,l}}\,\SU(p_{n,m}) \qquad \mbox{with} \quad
\sum_{(n,m)\in\quiver_{k,l}}\,(n+1)\,p_{n,m}\=p
\label{gaugegroupbreakSU}\eeq
where $(k+1)\,(l+1)$ is the number of elements of the weight set
$\quiver_{k,l}$.

The gauge coupling in $d$ dimensions should have mass dimension
$2-\frac d2$, and therefore we define $g^2=\tilde g^2/2(\pi\,R^2)^2$
as the $d$-dimensional gauge coupling constant. We then rescale
\beq
\phi_{n,m}^\pm~\longrightarrow~\mbox{$\frac{2g\,R}{\sqrt{n+1\pm1}}$}~
\phi_{n,m}^\pm \qquad \mbox{and} \qquad
A^{n,m}~\longrightarrow~\mbox{$\frac g{\sqrt{n+1}}$}~A^{n,m}
\label{phiArescale}\eeq
so that the scalar and gauge fields have the correct canonical
dimensions and kinetic term normalizations for a $d$-dimensional field
theory (with dimensionless coordinates). The Higgs potential in the
scalar fields $\phi^\pm_{n,m}$ in (\ref{SklYM}) generically leads to
spontaneous symmetry breaking, as a direct consequence of the
non-trivial background instanton and monopole charges on $\C
P^2$. Since this potential is a sum of non-negative terms, it is easy
to write down the general structure of the vacua in the Higgs sector
of the field theory. In particular, they obey the equations 
\bea
\phi_{n,m}^\pm{}^\dag\,\phi_{n,m}^\pm&=&
\frac{(n+1\pm1)\,\Lambda_{k,l}^\pm(n,m)^2}
{4g^2\,R^2}~\Idd_{p_{n,m}} \ , \nonumber \\[4pt] \phi_{n,m}^\pm\,
\phi_{n,m}^\pm{}^\dag&=&\frac{(n+1\pm1)\,\Lambda_{k,l}^\pm(n,m)^2}
{4g^2\,R^2}~\Idd_{p_{n\pm 1,m+3}} \ .
\label{Higgsvacgen}\eea
The vanishing of the last two terms in (\ref{SklYM}) represent the
relations of the quiver (\ref{Figquiver})~\cite{LPS3} and has a
natural algebraic meaning in terms of the operators
\bea
\mphi^\pm&:=&\sum_{(n,m)\in\quiver_{k,l}}\,
\Lambda_{k,l}^\pm(n,m)^{-1}\,\phi_{n,m}^\pm\otimes
\bigg(~\sum_{q\in\rel_n}\,\left(\,\sqrt{n\pm q+1\pm1}~
\big|\npmoverqpo\,,\,m+3\big\rangle\big\langle\noverq\,,\,m\big|
\right. \label{phipmgrconn} \\ && \hspace{7cm} \left.
+\,\sqrt{n\mp q+1\pm1}~\big|\npmoverqmo\,,\,m+3\big\rangle\big
\langle\noverq\,,\,m\big|\,\right)\,\bigg) \nonumber
\eea
defined with respect to the Biedenharn basis of
\S\ref{invgaugefields}. Then, in addition to (\ref{Higgsvacgen}), the
Higgs vacua are determined by the matrix commutativity relations
\beq
\big[\mphi^+\,,\,\mphi^-\big]\=0 \qquad \mbox{and}
\qquad \big[\mphi^+\,,\,\mphi^-\,^\dag\,\big]\=0 \  .
\label{Higgsvaccomms}\eeq

When $p_{n,m}=r$ for all weights $(n,m)\in\quiver_{k,l}$, the gauge
symmetry reduction is given by
\beq
\SU(p)~\longrightarrow~\uo^{(k+1)\,(l+1)-1}\times
\SU(r)^{(k+1)\,(l+1)} \qquad \mbox{with} \quad p\=r\,d^{k,l}
\label{SUpredalleq}\eeq
where $d^{k,l}$ are the dimensions (\ref{dimkl}). In this special case
an explicit solution of (\ref{Higgsvacgen}) is given by
$\phi_{n,m}^\pm=\phi_{n,m}^\pm{}^0$, where
\beq
\phi_{n,m}^\pm{}^0=\frac{\sqrt{n+1\pm1}~\Lambda_{k,l}^\pm(n,m)}
{2g\,R}~U^\pm_{n,m}
\label{phinmpmexplgen}\eeq
for $(n,m)\in\quiver_{k,l}$. This solution involves $2k\,l+k+l$
unitary degrees of freedom $U^\pm_{n,m}\in\urm(r)$, one for each Higgs
field $\phi^\pm_{n,m}$. We can associate each such unitary group
element with a link of the lattice (\ref{Figquiver}), which defines a
gauge field on the quiver lattice. However, they are not all
independent, because the commutation relations (\ref{Higgsvaccomms})
require that they obey
\beq
U^-_{n+1,m+3}\,U^+_{n,m}=U^+_{n-1,m+3}\,U^-_{n,m} \ ,
\label{Ucondition}\eeq
which is equivalent to requiring that their oriented product around
the four links of any plaquette in the quiver lattice must be equal to
unity. Thus the Higgs vacua correspond to \emph{flat} connections of
lattice gauge theory on the finite quiver lattice. However, there is
no vacuum moduli space, because we can set $k\,l+k+l$ of these unitary
degrees of freedom to the identity using a gauge transformation in the
$\urm(r)^{(k+1)\,(l+1)-1}$ subgroup of (\ref{SUpredalleq}), and then
eliminate the remaining ones using the $k\,l$ plaquette relations
(\ref{Ucondition}). Thus the solution (\ref{phinmpmexplgen}) 
breaks the gauge symmetry of
the $d$-dimensional field theory on $M$ to the diagonal $\SU(r)$
subgroup, leaving $(k\,l+k+l)\,r^2$ massive gauge bosons and
$(3k\,l+k+l)\,r^2$ physical Higgs fields. This mechanism induces
physical masses proportional to $\frac1R$. In subsequent sections we
shall work out some explicit examples.

\subsection{Reduction of the Dirac action\label{Yukawa}}

To describe the form of the fermionic action for the invariant spinor
fields constructed in \S\ref{InvSpinors}, we first need to set up some
Clifford algebra notation. The left-invariant one-forms defined in
(\ref{betacomps}) are proportional to orthonormal one-forms on $\CP^2$
and they define vierbeins $e^i{}_a $ on $\CP^2$ through
\beq
\beta^i\={\frac 1R}\,e^i{}_a ~\diff y^a \qquad \mbox{and} \qquad 
\betab^i\=\frac 1R\, e^{\bar\imath}{}_\ab~\diff\bar y^a \ ,
\label{betavier}\eeq
where $i=1,2$ is an orthonormal index and $a =1,2$ is a coordinate
index. With $M$ a complex manifold as in \S\ref{InvSpinors}, the
generators of the Clifford algebra $\cliff(M\times \C P^2)$ obey 
\beq
\Gamma^A\,\Gamma^B+\Gamma^B\,\Gamma^A\= 
-2\,\Gcal^{AB}~\Idd_{2^{d/2+2}}
\qquad\mbox{with}\quad A,B\=1,\dots,d+4 \ .
\label{2n2Cliffalg}\eeq
The gamma-matrices in (\ref{2n2Cliffalg}) may be decomposed as
\beq
\bigl\{\Gamma^A\bigr\}\=\bigl\{\Gamma^\mu,\Gamma^a ,
\Gamma^{\ab}\bigr\} \qquad\mbox{with}\quad \Gamma^\mu\=\gamma^\mu
\otimes\Idd_4 \ , ~~ \Gamma^a \=\gamma\otimes\tau^a 
\quad\mbox{and}\quad \Gamma^\ab \=\gamma\otimes\tau^\ab
\label{gamma2n2decomp}\eeq
where our convention is $\Gamma^a
\,\Gamma^\ab+\Gamma^\ab\,\Gamma^a=-\Gcal^{a \ab}~\Idd_{2^{d/2+2}}$ in
complex coordinates. The $2^{d/2}\times2^{d/2}$ matrices
$\gamma^\mu=-(\gamma^\mu)^\dag$ act locally on the twisted spinor
module $\underline{\Delta}\,(M)$ over the Clifford algebra bundle
$\cliff(M)\to M$ with the relations
\beq
\gamma^\mu\,\gamma^\nu+\gamma^\nu\,\gamma^\mu\=-2\,G^{\mu\nu}~
\Idd_{2^{d/2}}
\qquad\mbox{with}\quad \mu,\nu\=1,\dots,d \ ,
\label{2nCliffalg}\eeq
while
\beq
\gamma\=\frac{\im^{d/2}\,\sqrt{G}}{d!}~
\epsilon_{\mu_1\cdots\mu_{d}}\, 
\gamma^{\mu_1}\cdots\gamma^{\mu_{d}} \qquad\mbox{with}\quad
(\gamma)^2\=\Idd_{2^{d/2}} \quad\mbox{and}\quad
\gamma\,\gamma^\mu\=-\gamma^\mu\,\gamma
\label{chiralityop}\eeq
is the corresponding chirality operator. Here
$\epsilon_{\mu_1\dots\mu_{d}}$ is the Levi-Civita symbol with
$\epsilon_{12\cdots d}=+\,1$.

The coordinate basis gamma-matrices $\tau^a $ and $\tau^\ab$ on
$\CP^2$ are related to their orthonormal counterparts by
\beq 
\sigma^i\=\im e^i{}_a\,\tau^a \qquad \mbox{and} \qquad
\sigma^{\bar\imath}\=-\im e^{\bar\imath}{}_\ab\, \tau^\ab
\label{gammavier}\eeq
with the normalisation chosen so that
\beq 
\sigma^i\,\sigma^{\bar\jmath}+\sigma^{\bar\jmath}\,
\sigma^i=\delta^{ij}\Idd_4 \ .
\label{CliffAlgebra}
\eeq
It is a standard construction~\cite{GSW} to choose a basis in which
$(\sigma^i)^2=(\sigma^{\bar\imath}\,)^2 =0$, and to associate
$\sigma^{\bar\imath}$ and $\sigma^i$ respectively with creation and
annihilation operators acting on a fermionic Fock space with vacuum
vector $|\Omega\rangle$ such that $\sigma^i|\Omega\rangle=0$. A
general Fock space state
\beq 
|\chi\rangle:=\chi_0(y,\bar y)\otimes
|\Omega\rangle + \chi_{\ib}(y,\bar y) \otimes
\sigma^{\ib} |\Omega\rangle +\mbox{$\frac 12$}\,
\chi_{\ib\,\jb}(y,\bar y) \otimes\sigma^{\ib\,\jb} |\Omega\rangle \ ,
\label{FockFermion}
\eeq
with $\sigma^{\ib\,\jb}:=\frac12\,[\sigma^\ib,\sigma^\jb\,]$,
corresponds locally to a Dirac spinor on $\CP^2$, though of course it 
may not extend to a global spinor field. The chirality operator on
$\CP^2$ is $\widetilde\sigma= \big[\sigma^1,\sigma^{\bar
  1}\,\big]\,\big[\sigma^2,\sigma^{\bar2}\,\big]$ and $\chi_\ib$ are
the two components of a negative chirality spinor, while $\chi_0$ and
$\chi_{\bar 1\,\bar 2}$ are the two components of a positive chirality
spinor. In terms of holonomy, $\chi_\ib$ is a doublet of $\SU(2)$
while $\chi_0$ and $\chi_{\bar 1\,\bar 2}$ are both $\su$-singlets. 

An alternative way of understanding these representation assignments,
which will be useful in later sections, follows from the general
construction in~\cite{BDHom}. Spinor fields on $\CP^2$ transform in
the $4\times 4$ (reducible) spinor representation of
$H=\SU(2)\times\uo$ given by
\bea
\Sigma_{E_{\alpha_1}}\=\sigma^1\,\sigma^{\bar 2}\qquad &\mbox{and}&
\qquad \Sigma_{E_{-\alpha_1}}\=\Sigma_{E_{\alpha_1}}{}^\dag
\=\sigma^2\,\sigma^{\bar 1} \ , \nonumber\\[4pt] 
\Sigma_{H_{\alpha_1}}\=\sigma^1\,\sigma^{\bar 1} -
\sigma^2\,\sigma^{\bar 2} \qquad &\mbox{and}& \qquad
\Sigma_{H_{\alpha_2}}\=\sigma^1\,\sigma^{\bar 1} +
\sigma^2\,\sigma^{\bar 2}-{\bf 1}_4 \ .
\label{SpinorRep}\eea
These generators constitute a traceless representation of the ${\rm
  su}(2)\oplus\uoL$ subalgebra of (\ref{HEEcomms}), as is easily
checked using (\ref{CliffAlgebra}). The representation content is
revealed by evaluating the second order Casimir invariants of $\SU(2)$
and $\urm(1)$ to get
\beq
\mbox{$\frac 1 2$}\, \bigl| \Sigma_{E_{\alpha_1}}\bigr|^2
+\mbox{$\frac 1 2$}\,\big(\Sigma_{H_{\alpha_1}}\big)^2 \= \mbox{$\frac 
  38$}\,\big(\Idd_4-\widetilde\sigma\,\big) \qquad \mbox{and} \qquad
\mbox{$\frac 1 2$}\,\big(\Sigma_{H_{\alpha_2}}\big)^2\= \mbox{$\frac
  18$}\,\big(\Idd_4+\widetilde\sigma\,\big) \ .
\eeq
It follows that negative chirality spinors live in the representation
$\underline{(1,0)}$, while positive chirality spinors are given by a
pair of $\SU(2)$-singlets with opposite hypercharge $Y=\pm\,1$ in the
$H$-module $\underline{(0,\pm\,3)}$.

These states correspond respectively to the instanton bundle $\Ical$,
with fibres transforming under the representation $\underline{(1,0)}$,
and the monopole line bundles $\Lcal_{\pm\,3/ 2}$, with fibres
transforming under the representation $\underline{(0,\pm\, 3)}$. None
of these bundles is globally well-defined of course. In order to get
well-defined bundles on $\CP^2$, we must tensor the would-be spin
bundle with non-trivial gauge bundles $\Ical_n\otimes\Lcal_{\widetilde
  m/2}$ whose fibres transform according to the representation
$\underline{(n,\widetilde m\,)}$ of $\SU(2)\times\urm(1)$ with $n$ and
$\widetilde m=m+c$ of opposite even/odd integer parity. These bundles
do not exist on their own but, as described in \S\ref{InvSpinors},
their tensor product does. The complete $\SU(2)\times\urm(1)$
representation content of these bundles is given by the decomposition
into irreducible modules
\bea
\underline{(n,\widetilde m\,)}\,
\otimes \,\big(\,\big[\,\underline{(1,0)}\,\big]~
\oplus~\big[\,\underline{(0,3)}\,\oplus\,\underline{(0,-3)}\,\big]\,
\big) &=& \big[\,\underline{(n+1,\widetilde m\,)}\,
\oplus\,\underline{(n-1,\widetilde m\,)}\,\big] \nonumber\\ &&
\,\oplus~\big[\,\underline{(n,\widetilde m+3)}\,\oplus\,
\underline{(n,\widetilde m-3)}\,\big]
\label{SpinBundleDecomposition}\eea
where the square brackets segregate the spinor chiralities.

For each weight $(n,m)\in\quiver_{k,l}$, the complete spectrum of the
twisted Dirac operator on $\CP^2$ consists of $4(n+1)$ families of
infinite discrete sequences of eigenvalues, one family for each state
on the right-hand side of (\ref{SpinBundleDecomposition}). The
non-zero eigenvalues come in positive and negative pairs giving the
$2(n+1)$ sequences listed below. The
spectrum therefore grows rapidly more complicated as $n$
increases. Note that at least some of the corresponding eigenspinors
must necessarily have different assignments of $\SU(2)\times\urm(1)$
quantum numbers for their two chiral components. After dimensional
reduction, the eigenspinors on $\CP^2$ with non-zero eigenvalues will
induce a total of $4d^{k,l}$ infinite discrete families of fermion
fields on $M$ with both a kinetic mass term, given by the Dirac
eigenvalues on $\CP^2$, and Yukawa couplings.

The eigenvalues and their multiplicities can be read off from the
explicit formulas of~\cite[\S B]{BDHom}. There are $2n+2$ infinite
sequences, with $n+2$ families coming from the states in
$\underline{(n',\widetilde m\,)}$ with $n'=n+1$
and $n$ families coming from the states with $n'=n-1$. Denoting the
eigenvalues by ${\lambda_N\over R}$ and their degeneracies by $d_N$,
we distinguish each sequence by $n+2$ integers $\eta_+$ and $n$
integers $\eta_-$ with
\bea\label{DiracEigenvalues1}
\lambda_N^2&=&N\,(N+n+3)+\eta_+^2 +\frac{|\eta_+|}2\,\big(2N+n+3
\big)-\frac {\eta_+} 2 \,\big|\,\widetilde m\,\big| + n + 2 \ ,
 \\[4pt]
d_N&=& \frac 1 8 \,\Big(2N+n+3 + \epsilon_+\, \big(4\eta_+ -
|\,\widetilde m\,|\big)\Bigr)\,
\Big(2N+n+3 -\epsilon_+ \, \big(2\eta_+ -|\,\widetilde m\,|\big)\Big)\, 
\Big(2N+n+3 + |\eta_+|\Big)\nonumber
\eea
for $n'=n+1$, and
\bea\label{DiracEigenvalues2}
\lambda_N^2&=&N\,(N+n+1)+\eta_-^2 +\frac{|\eta_-|} 2\,\big(2N+n+1
\big)- \frac {\eta_-} 2 \,\big|\,\widetilde m\,\big| \ ,
\\[4pt]
d_N&=& \frac 1 8\, \Big(2N+n+1 + \epsilon_-\, \big(4\eta_--
|\,\widetilde m\,|\big)\Big)\,
\Big(2N+n+1 -\epsilon_- \, \big(2\eta_- -|\,\widetilde
m\,|\big)\Big)\, \Big(2N+n+1 + |\eta_-|\Big) \nonumber
\eea
for $n'=n-1$, where
\beq
\eta_\pm\=-\mbox{$\frac12$}\,\big(n\pm 1+|\,\widetilde m\,|\big)
,\ldots, \mbox{$\frac12$}\, \big(n\pm 1 +|\,\widetilde m\,|\big) \ .
\eeq
In both cases $N=0,1,\ldots$, while $\epsilon_\pm=1$ for $\eta_\pm\ge
0$ and $\epsilon_\pm=-1$ for $\eta_\pm< 0$. We shall see some explicit
examples in the following sections.

We will now construct the $\Ecal^{k,l}$-twisted Dirac operator
$\Diraccal=\Gamma^A\,{\cal D}_A$ on $\man=M\times\C P^2$,
corresponding to the equivariant gauge potential $\ca$ in
(\ref{calACP2Cklrep}) and acting on the spinor fields
(\ref{Psiequiv}), in terms of the spin$^c$ Dirac operator
(\ref{DiracCP2}) on $\CP^2$ and the $E^{k,l}$-twisted spin$^c$ Dirac
operator $\Dirac=\gamma^\mu\,D_\mu$ on $M$. The latter operator is
given by
\beq
\Dirac=\sum_{(n,m)\in\quiver_{k,l}}\,\big( \pa_M-\mbox{$\frac12$}\;
\kappa \kern -6pt/\;+A^{n,m}\kern -22pt/ \kern+16pt
\big)\otimes\pi_{n,m}
\label{DiracM}\eeq
where $\pa_M$ is the naive Dirac operator on $M$ involving only the
spin connection on the principal ${\rm SO}(d)$-bundle $P_{{\rm
    SO}(d)}\to M$ and the generators of ${\rm SO}(d)$ in the spinor
representation, while $\kappa$ is an anti-hermitean connection on the
canonical line bundle $K\to M$. Using (\ref{betavier}) and
(\ref{gammavier}) one then finds
\bea
\Diraccal&=&\Dirac\otimes\Idd_4+\gamma\otimes\Dirac_{\C P^2}
\label{Diracgradeddef}\\ && +\,\frac1R~\sum_{(n,m)\in\quiver_{k,l}}\,
\Big(\phi_{n,m}^+\,\gamma\otimes\sigma^+_{n,m}+\phi_{n,m}^-\,
\gamma\otimes\sigma_{n,m}^--
{\phi_{n,m}^+}^\dag\,\gamma\otimes\sigma^+_{n,m}{}^\dag-
{\phi_{n,m}^-}^\dag\,\gamma\otimes\sigma_{n,m}^-{}^\dag\Big) \nonumber
\eea
where, in complete analogy with (\ref{betapmdef}) and
(\ref{betanmdef}), we have defined
\beq 
\sigma^\pm \= \im\sigma^{\bar 1}\otimes E^\pm_{\alpha_1+\alpha_2}+\im
\sigma^{\bar 2}\otimes E^\pm_{\alpha_2}\=
\sum_{(n,m)\in\quiver_{k,l}}\,\sigma^\pm_{n,m} \ .
\label{sigmapmdef}\eeq

Using the twisted Dirac operator (\ref{Diracgradeddef}), we may define
an euclidean fermionic action functional on the space of ${\rm
  L}^2$-sections (\ref{Psiequiv}) by
\beq
S^{k,l}_{\rm D}:=\int_{M\times\C P^2}\,\diff^{d+4}x~\sqrt {|\Gcal|}~
\Psi^\dag\,\Diraccal\Psi \ , 
\label{EDdef}\eeq
where $\Psi$ has canonical mass dimension $\frac12\,(d+3)$. In
lorentzian signature the adjoint spinor $\Psi^\dag$ should be replaced
by $\overline\Psi:=\frac 1{\sqrt{-\Gcal^{00}}}\,\Psi^\dag\Gamma^0$.
For definiteness, we shall only consider the case where the spinor
field $\Psi$ transforms under the fundamental representation of the
initial gauge group $\SU(p)$. Other fermion representations of
$\SU(p)$ can be treated similarly. We substitute (\ref{Psipsichi}) and
integrate over $\CP^2$ in (\ref{EDdef}). The zero modes of
$\Dirac_{\CP^2}$ can be chosen to be orthogonal and normalised such
that
\beq
\int_{\CP^2}\,\chi^\pm_{n',m';\ell'}{}^\dag\;\chi^\pm_{n,m;\ell}~
\beta_{\rm vol} \=\delta_{n,n'}\,\delta_{m,m'}\,\delta_{\ell,\ell'} 
\qquad \mbox{and} \qquad
\int_{\CP^2}\,\chi^\mp_{n',m';\ell'}{}^\dag\;\chi^\pm_{n,m;\ell}~
\beta_{\rm vol}\=0 \ ,
\label{0modenorm}\eeq
where the second equality follows from the fact that the sets
$\quiver_{k,l}^+$ and $\quiver_{k,l}^-$ in (\ref{weightklpm}) are
disjoint. Since $\chi^\pm_{n,m;\ell}$ are spinor harmonics on $\CP^2$,
one might naively expect that the fermion fields $\psi_{n,m;\ell}$ and
$\widetilde\psi_{n,m;\ell}$ will be massless spinors on $M$. However,
the Higgs field terms in (\ref{Diracgradeddef}) can give rise to
Yukawa couplings and, due to spontaneous symmetry breaking, induce
masses of order $\frac 1R$ to the $d$-dimensional spinors. We shall
now explain precisely how this comes about.

Recall that the fermion zero modes depend on the twisting parameter
$c=2\,\ct$ introduced in \S\ref{InvSpinors}. We will now show how to
\emph{uniquely} fix this free parameter such that the reduction of the
action (\ref{EDdef}) generically contains Yukawa couplings. We
consider background gauge fields on $\CP^2$ for which the index
(\ref{nunm}) takes values $\nu_{n,m}=\pm\,1$. The spinor harmonic
modes are particularly simple in this case~\cite{BDCP2}. They arise as
a result of the gauge connections of the $\su\times\uo$ gauge theory
on $\CP^2$ exactly cancelling the spin connection, so that the Dirac
operator truncates to the (untwisted) Dolbeault operator on $\CP^2$
and the components of the spinors in (\ref{FockFermion}) are simply
constants. Note that this can only occur when $n=0,1$, and hence by
(\ref{mnqjpm}) for $m=-2(k-l),-2(k-l)\pm3$. For any given irreducible
$\sut$-representation $\underline{C}^{k,l}$, it is easy to deduce from
(\ref{nunm}) that the unique spin$^c$ structure on $\CP^2$
accommodating these fields has twisting parameter
\beq
c=2(k-l)-3 \ .
\label{twistunique}\eeq
Then the chiral fermion mode with $(n,m)=(0,-2(k-l))$ (and
$\nu_{n,m}=+1$) will have a Yukawa coupling to the antichiral mode
with $(n,m)=(1,-2(k-l)+3)$ (and $\nu_{n,m}=-1$).\footnote{For
  $l\geq1$, one can alternatively choose $c=2(k-l)+3$, and couple the
  chiral mode with $(n,m)=(0,-2(k-l))$ to the antichiral mode with
  $(n,m)=(1,-2(k-l)-3)$.}

The positive chirality mode with respect to the Biedenharn basis of
\S\ref{invgaugefields} is given by
\beq
\chi^+_{0,-2(k-l)}=|\Omega\rangle\otimes
\big|{\stackrel{\scriptstyle 0}
{\scriptstyle 0}}\,,\,-2(k-l)\big\rangle
\label{poschirmode}\eeq
while the negative chirality mode, which is a doublet of the $\SU(2)$
gauge theory on $\CP^2$, is
\beq
\chi^-_{1,-2(k-l)+3}=\frac1{\sqrt2}\,\Big(\sigma^{\bar 1}
|\Omega\rangle\otimes\big|{\stackrel{\scriptstyle 1}
{\scriptstyle 1}}\,,\,-2(k-l)+3\big\rangle+\sigma^{\bar 2}
|\Omega\rangle\otimes\big|{\stackrel{\scriptstyle 1}
{\scriptstyle -1}}\,,\,-2(k-l)+3\big\rangle\Big) \ .
\label{negchirmode}\eeq
From the explicit formulas (\ref{lambdaklnm}) one finds
$\Lambda_{k,l}^-(0,-2(k-l))= \Lambda_{k,l}^-(1,-2(k-l)+3)=0$, and
consequently the only contributing operator from (\ref{sigmapmdef})
is given by
\beq
\sigma^+_{0,-2(k-l)}=\mbox{$\frac\im{\sqrt2}$}\,\sqrt{k\,(l+2)}\,
\Big(\sigma^{\bar 1}\otimes\big|{\stackrel{\scriptstyle 1}
{\scriptstyle 1}}\,,\,-2(k-l)+3\big\rangle
\big\langle{\stackrel{\scriptstyle 0}
{\scriptstyle 0}}\,,\,-2(k-l)\big|+\sigma^{\bar 2}\otimes
\big|{\stackrel{\scriptstyle 1}
{\scriptstyle -1}}\,,\,-2(k-l)+3\big\rangle
\big\langle{\stackrel{\scriptstyle 0}
{\scriptstyle 0}}\,,\,-2(k-l)\big|\Big)
\label{sigma0mode}\eeq
with
\bea
\sigma^+_{0,-2(k-l)}\chi^+_{0,-2(k-l)}&=&
\im\sqrt{k\,(l+2)}~\chi^-_{1,-2(k-l)+3} \ ,
\nonumber \\[4pt] \sigma^+_{0,-2(k-l)}\chi^-_{1,-2(k-l)+3}&=&
\im\sqrt{k\,(l+2)}~\chi^+_{0,-2(k-l)} \ .
\label{sigmachiaction}\eea
These are then the only surviving contributions from the Higgs field
terms in (\ref{Diracgradeddef}) after integration over $\CP^2$ using
(\ref{0modenorm}).

We now rescale the bosonic fields as in (\ref{phiArescale}) and the
fermionic fields as
\beq
\psi_{n,m;\ell}~\longrightarrow~\mbox{$\frac1{\sqrt2\,\pi\,R^2}$}\,
\psi_{n,m;\ell} \qquad \mbox{and} \qquad
\widetilde\psi_{n,m;\ell}~\longrightarrow~
\mbox{$\frac1{\sqrt2\,\pi\,R^2}$}\,\widetilde\psi_{n,m;\ell} \ ,
\label{fermrescale}\eeq
in order to give all fields the correct canonical dimensions and
kinetic term normalizations on $M$. Putting everything together, the
dimensional reduction of the Dirac action (\ref{EDdef}) is given by
\bea
S_{\rm D}^{k,l}&=&\int_{M}\,\diff^dx~\sqrt{|G|}~\bigg[~
\sum_{(n,m)\in\quiver_{k,l}^+}~
\sum_{\ell=1}^{\nu_{n,m}}~\big(\psi_{n,m;\ell}
\big)^\dag\,\Dirac\bigl(\psi_{n,m;\ell}\bigr) \nonumber\\ &&
\hspace{3cm} +\,\sum_{(n,m)\in\quiver_{k,l}^-}~
\sum_{\ell=1}^{|\nu_{n,m}|}~
\big(\widetilde\psi_{n,m;\ell}\big)^\dag\,
\Dirac\bigl(\widetilde\psi_{n,m;\ell}\bigr) \label{EDdimred}\\
&&\hspace{3cm} +\,\sqrt{2k\,(l+2)}\,g\,
\Big(\left(\psi_{0,-2(k-l)}\right)^\dag \,
\phi^+_{0,-2(k-l)}{}^\dag\,\psi_{1,-2(k-l)+3} \nonumber\\ &&
\hspace{6cm} +\,\big(\psi_{1,-2(k-l)+3}\big)^\dag\,
\phi^+_{0,-2(k-l)}\,\psi_{0,-2(k-l)}\Big)\bigg] \ , \nonumber
\eea
where we have abbreviated $\psi_{0,-2(k-l)}:=\psi_{0,-2(k-l);0}$ and
$\psi_{1,-2(k-l)+3}:= \gamma\widetilde\psi_{1,-2(k-l)+3;0}$. The fermion
fields $\psi_{n,m;\ell}$ and $\widetilde\psi_{n,m;\ell}$ for each
$\ell=1,\dots,|\nu_{n,m}|$ transform in the fundamental representation
of $\SU(p_{n,m})$. The dimensionally reduced field theory thus
contains Yukawa interactions for all $k>0$. If the Higgs field
$\phi^+_{0,-2(k-l)}$ acquires a non-zero vacuum expectation value
$\phi^+_{0,-2(k-l)}{}^0$ by dynamical symmetry breaking, then the
fermion fields $\psi_{0,-2(k-l)}$ and $\psi_{1,-2(k-l)+3}$ acquire a
mass matrix. In the special case (\ref{phinmpmexplgen}), the positive
eigenvalue of this mass matrix is
\beq
\mu_{k,l}=\frac{k\,(l+2)}{\sqrt2\,R} \ .
\label{mukl}\eeq

\subsection{Chain reductions\label{YMHchains}}

To exemplify the quantitative differences between the quiver
gauge theory defined by (\ref{SklYM}) and those studied
in~\cite{Dolan09} which are obtained via $\su$-equivariant dimensional
reduction over the projective line $\C P^1$, let us set $l=0$ and
consider the reductions associated to the irreducible
$\sut$-representations $\underline{C}^{k,0}$. In this case
$j_-=0$ in (\ref{mnqjpm}), so that the monopole charges and instanton
ranks are correlated as $(n,m)=(n,3n-2k)$ with
$n=0,1,\dots,k$. With $p_n:=p_{n,3n-2k}$, the explicit gauge symmetry
breaking pattern is given in this limit by
\beq
\SU(p)~\longrightarrow~\uo^k\times\prod_{n=0}^k\,\SU(p_n) \qquad
\mbox{with} \quad \sum_{n=0}^k\,(n+1)\,p_n\=p \ .
\label{explsymbrl0}\eeq
Although similar to the symmetry reduction patterns of~\cite{Dolan09},
the rank decompositions in (\ref{explsymbrl0}) are controlled
explicitly by the instanton ranks $n+1$.

From (\ref{lambdaklnm}) one also finds $\Lambda_{k,0}^-(n,3n-2k)=0$
and $\Lambda_{k,0}^+(n,3n-2k)=\sqrt{(n+1)\,(k-n)}$. It follows that
$\betab_{n,3n-2k}^-=0$ for all $n=0,1,\dots,k$, and consequently all
fields $\phi_{n,m}^-$ are absent from (\ref{calACP2Cklrep}). Thus in
this case the two-dimensional quiver lattice of equivariant fields on
$M$ labelled by $\quiver_{k,l}$ truncates to a one-dimensional
\emph{chain}
\beq
\mbox{\includegraphics[width=2.5in]{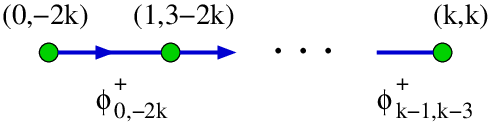}}
\label{Figchain}\eeq
Denote $\phi_{n+1}:=\phi^+_{n,3n-2k}$ and $A^n:=A^{n,3n-2k}$, with
\beq
F^n\=\diff A^n+\mbox{$\frac g{\sqrt{n+1}}$}\,A^n\wedge A^n \qquad
\mbox{and} \qquad D\phi_{n+1}\=\diff\phi_{n+1}+
g\,\big(\mbox{$\frac1{\sqrt{n+2}}$}\,A^{n+1}\,\phi_{n+1}-
\mbox{$\frac1{\sqrt{n+1}}$}\,\phi_{n+1}\,A^n\big) \ .
\label{FDphichain}\eeq
Then the action (\ref{SklYM}) reduces to
\bea
S_{\rm YM}^{k,0}&=&\int_{M}\,\diff^{d}x~\sqrt{|G|}\ \bigg[\,
\sum_{n=0}^k\,\tr^{~}_{p_n\times p_n}\Big(\mbox{$\frac14$}\,
\bigl(F_{{\mu}{\nu}}^n\bigr)^\+\,\bigl(F^{n\,{\mu}{\nu}}\bigr)
+ \bigl(D_{{\mu}} \phi_{n+1}\bigr)
\,\bigl(D^{{\mu}} \phi_{n+1}\bigr)^\+ \nonumber\\ &&
\qquad\qquad\qquad\qquad\qquad\qquad\qquad +\,\bigl(D_{{\mu}}
\phi_{n}\bigr)^\+\, \bigl(D^{{\mu}}\phi_{n}\bigr)\Big)+
V(\phi_1,\dots,\phi_k)\,\bigg]
\label{Sk0YM}\eea
with $\phi_0=\phi_{k+1}=0$ and the Higgs potential
\bea
V(\phi_1,\dots,\phi_k)&=&2g^2\,\sum_{n=0}^k\,
\tr^{~}_{p_n\times p_n}\bigg[\,\frac1{n+2}\,\Big(
\frac{(n+1)\,(n+2)\,(k-n)}{4g^2\,R^2}~\Idd_{p_n}-\phi_{n+1}^\dag\,
\phi_{n+1}\Big)^2 \nonumber \\ && \qquad\qquad\qquad\qquad 
+\,\frac1{n}\,\Big(
\frac{n\,(n+1)\,(k-n+1)}{4g^2\,R^2}~\Idd_{p_n}-\phi_n\,
\phi_n^\dag\Big)^2\,\bigg] \ .
\label{Higgspotchain}\eea
This potential is minimized by scalar field configurations $\phi_n$
obeying
\beq
\phi_n^\dag\,\phi_n\=\frac{n\,(n+1)\,(k-n+1)}{4g^2\,R^2}~\Idd_{p_{n-1}}
\qquad \mbox{and} \qquad
\phi_n\,\phi_n^\dag\=
\frac{n\,(n+1)\,(k-n+1)}{4g^2\,R^2}~\Idd_{p_{n}} \ .
\label{Higgsvacchain}\eeq

In the special case where $p_0=p_1=\cdots=p_k=r$, so that the gauge
symmetry reduction is given by
\beq
\SU(p)~\longrightarrow~\uo^k\times\SU(r)^{k+1} \qquad \mbox{with}
\quad p\=\mbox{$\frac12$}\,r\,(k+1)\,(k+2) \ ,
\label{gaugebrchain}\eeq
an explicit solution of (\ref{Higgsvacchain}) is given by
$\phi_n=\phi_n^0$, where
\beq
\phi_n^0=\frac1{2g\,R}~\sqrt{n\,(n+1)\,(k-n+1)}~U_n
\label{phin0chain}\eeq
for $n=1,\dots,k$. The independent unitary degrees of freedom
$U_n\in{\rm U}(r)$ can be removed using a $\uo^k\times\SU(r)^k$ gauge
transformation, and this solution breaks the gauge symmetry to
$\SU(r)$. There are $k\,r^2$ massive gauge bosons, and $k\,r^2$
physical Higgs fields represented by $r\times r$ hermitean matrices
$h_n$, $n=1,\dots,k$ with $\phi_n=\phi_n^0+h_n$. The Higgs and vector
boson masses, both proportional to $\frac1R$, can be worked out by
substitution into the action (\ref{Sk0YM}). A completely analogous
analysis follows in the cases with $k=0$, though there will be
quantitative differences. While the physics of the
dynamical symmetry breaking for these systems is qualitatively
analogous to the cases studied in~\cite{Dolan09}, the quantitative
features are significantly different due to the different forms of the
interactions in (\ref{FDphichain}) and of the Higgs potential
in~(\ref{Higgspotchain}). These differences are due to the fact that
while only monopole backgrounds on $\CP^1$ contribute to the
equivariant dimensional reduction considered in~\cite{Dolan09}, here
both instanton and monopole charges on $\CP^2$ affect the quiver gauge
theory.

The quantitative differences from the $\CP^1$ models are somewhat more
drastic in the fermionic sector, due to the large asymmetry between
the positive and negative chirality spinor harmonics
on $\CP^2$ in the limit $l=0$. With the spin$^c$ twist
(\ref{twistunique}), the index (\ref{nunm}) in this limit becomes
\beq
\nu_n~:=~\nu_{n,3n-2k}\=\mbox{$\frac12$}\,(n+1)\,(n-2)\,(2n-1)
\label{nunl0}\eeq
for $n=0,1,\dots,k$. Thus there is only a single antichiral mode
$\widetilde\psi:= \widetilde\psi_{1,3-2k;0}$, whose chiral partner is
$\psi:=\psi_{0,-2k;0}$. The remaining fermion fields
$\psi_{n;\ell}:=\psi_{n,3n-2k;\ell}$ on $M$ for $n>2$ are all induced
from positive chirality spinor harmonics on $\CP^2$, transform in the
fundamental representation of $\SU(p_n)$, and have gauge interactions
given by
\beq
\Dirac\psi_{n;\ell}=\big(\pa_M-\mbox{$\frac12$}\;
\kappa \kern -6pt/\;+\mbox{$\frac g{\sqrt{n+1}}$}\,
A^{n}\kern -12pt/ \kern+6pt \big)\psi_{n;\ell}
\label{Diracpsinell}\eeq
for each $\ell=1,\dots,\nu_n$. The fermionic action (\ref{EDdimred})
thereby truncates to
\beq
S_{\rm D}^{k,0}=\int_M\,\diff^dx~\sqrt{|G|}\ \bigg[\psi^\dag\,
\Dirac\psi+\widetilde\psi\,^\dag\,\Dirac\widetilde\psi+2\,\sqrt k\,g\,
\big(\psi^\dag\,\phi_1^\dag\,\gamma\widetilde\psi+
\widetilde\psi\,^\dag\,\phi_1\,\gamma\psi\big)+
\sum_{n=3}^k~\sum_{\ell=1}^{\nu_n}\,\psi_{n;\ell}^\dag\,\Dirac
\psi_{n;\ell} \bigg] \ ,
\label{SDk0}\eeq
and the fermion mass induced by the Higgs vacuum (\ref{phin0chain})
and the Yukawa interaction in (\ref{SDk0}) is
\beq
\mu_{k,0}=\frac{\sqrt2\,k}R \ .
\label{muk0}\eeq
In contrast to the Dirac-Higgs chains which arise from dimensional
reduction over $\CP^1$~\cite{Dolan09}, Yukawa interactions here exist
for all values of $k>0$. On the other hand, there are no Yukawa
interactions in (\ref{EDdimred}) in the limit $k=0$.

Furthermore, the construction of massive eigenspinors discussed in
\S\ref{Yukawa} proceeds by substituting $\widetilde m=3(n-1)$ for
$n=0,1,\dots,k$ in (\ref{SpinBundleDecomposition}) and involves
$4(n+1)$ families of states in multi-dimensional irreducible
representations of the $\SU(2)$ isospin group. This contrasts markedly
with the situation for spinors on $\CP^1$ where all irreducible
representations of the $\urm(1)$ holonomy group are one-dimensional
and spinors are two-component fields, so only two families of
eigenvalues ever arise from a single irreducible representation of the
$\urm(1)$ gauge group on $\CP^1$. These two families actually
correspond to a single family with equally paired positive and
negative eigenvalues.

\bigskip

\section{Dynamical symmetry breaking from the fundamental
  representation\label{Fundamental}}

\noindent
In this section we will work out the details of dynamical symmetry
breaking in the quiver gauge theory which is induced by
dimensional reduction from the three-dimensional fundamental
representation $\underline{C}^{1,0}$ of $\sut$. It is obtained by
setting $k=1$ in the class of models studied in \S\ref{YMHchains}.
The analysis in this case is completely analogous to that of the
fundamental $\su$ representations in the $\CP^1$ models
of~\cite{Dolan09}. We will determine the physical particle spectrum
and masses in several explicit instances, including symmetry
hierarchies which entail dynamical electroweak symmetry breaking.

\subsection{Spontaneous symmetry breaking}

For $k=1$, $l=0$ there are two weights in $\quiver_{1,0}$, with
$(n,m)=(1,1)$ and $(n,m)=(0,-2)$, and a single Higgs field
$\phi:=\phi_1=\phi_{0,-2}^+$ which is a $p_1\times p_0$ complex
matrix. The quiver lattice is simply a chain consisting of one link
\beq
\mbox{\includegraphics[width=1.2in]{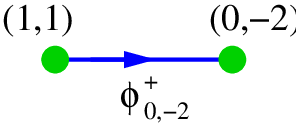}}
\label{Figfundamental}\eeq
Suppose that $p_1\ge p_0$. Then, with a suitable gauge choice,
the Higgs minimum can be put in the form
\beq
\phi^0=\frac 1{\sqrt 2\, g\,R}\,\begin{pmatrix} {\bf
    0}_{(p_1-p_0)\times p_0} \\ {\bf 1}_{p_0} \end{pmatrix} \ ,
\label{C10vacHiggs}\eeq
where ${\bf 0}_{(p_1-p_0)\times p_0}$ is a $(p_1-p_0)\times p_0$
matrix of zeroes. The gauge symmetry breaking sequence is given by
\beq
\SU(p)~\longrightarrow~\SU(p_0)\times\SU(p_1)\times\urm(1)~
\longrightarrow~\SU(p_1-p_0)\times\SU(p_0)_{\rm diag}\times\urm(1)' 
\qquad \mbox{with} \quad p\=p_0+2p_1 \ ,
\label{C10SUpbreakgen}\eeq
where the last step is dynamical symmetry breaking with $\SU(p_0)_{\rm
  diag}$ the diagonal $\SU(p_0)$ subgroup leaving ${\bf 1}_{p_0}$
invariant, and $\urm(1)'$ acts from the left on the top $p_1-p_0$ rows
of $\phi^0$. The case $p_0>p_1$ can be treated similarly. 

The gauge boson masses can be determined from the bicovariant
derivative in (\ref{FDphichain}), which in this case reads
\beq
D\phi=\diff\phi +g\,\big(\mbox{$\frac 1 {\sqrt 2}$}\,A^1\,
\phi -\phi\,A^0\big) \ .
\label{eq:DPhi}\eeq
For the moment we shall take the gauge potential $A^0$ to lie in ${\rm
  u}(p_0)$ and $A^1$ in $\urmL(p_1)$, as an overall $\urmL(1)$ part
will drop out. Let 
\beq 
A^1\=A^a_L\, \Bigl(\frac {\im\lambda_a}2\Bigr)+B_L\,
\frac \im {\sqrt{2 p_1}}~{\bf 1}_{p_1} \qquad \mbox{and} 
\qquad A^0\=A^{\tilde a}_R\, \Bigl(\frac {\im\lambda_{\tilde a}}2
\Bigr)+B_R\, \frac \im{\sqrt{2 p_0}}~{\bf 1}_{p_0} \ ,
\eeq
where $\lambda_a$ are Gell-Mann matrices for $\SU(p_0)$ with
$\tr_{p_0\times p_0}(\lambda_a\,\lambda_b)=2\delta_{ab}$, 
$\lambda_{\tilde a}$ are Gell-Mann matrices for $\SU(p_1)$, and the
square root factors are chosen so that the $\urm(1)$ generators have
the same normalisation as the Gell-Mann matrices. Then only the
combination  
\beq 
B:=\frac 1 {\sqrt p}\left(\sqrt {p_0} \;B_L - \sqrt{2p_1}\,B_R\right)
\eeq
appears in (\ref{eq:DPhi}), since the orthogonal combination 
$\frac 1 {\sqrt p}\left(\sqrt{2p_1}\, B_L + \sqrt{p_0}\;B_R\right)$
decouples as it should. With this notation, the bicovariant derivative
(\ref{eq:DPhi}) now reads
\beq
D\phi=\diff\phi+\mbox{$\frac {\im g} 2$}
\,\left(\mbox{$\frac 1{\sqrt{2}}$}\, A_L^a\, 
\lambda_a\,\phi - A^{\tilde a}_R \,\phi\, \lambda_{\tilde a}  +
\sqrt{\mbox{$\frac {p} {p_1\,p_0}$}}~B\,\phi\right) \ ,
\label{eq:Dphi}\eeq
from which we can obtain the gauge boson mass matrix $\mbf M$ 
by substituting the vacuum expectation value (\ref{C10vacHiggs}) of
the Higgs field to get
\beq 
\mbox{$\frac 12$}\,{\mbf A}^\top\,{\mbf M}^2\,
{\mbf A}=\tr_{p_0\times
  p_0}\Bigl(\big(D\phi^0\big)^\dag\,D\phi^0\Bigr)
\label{eq:AMA}\eeq
where $\mbf A$ is a column vector consisting of the vector bosons in
(\ref{eq:Dphi}). We will now work through some explicit examples.

\bigskip

\noindent
{\bf $\mbf{p_0=p_1=r}$ . \ } 
In this case one has
\beq 
\phi^0=\frac 1{\sqrt 2\,g\,R}~{\bf 1}_r
\label{HiggsVEV10}\eeq
and $\lambda_a=\lambda_{\tilde a}$. The symmetry breaking
pattern is
\beq
\SU(3r)~\longrightarrow~\SU(r)\times\SU(r)_{\rm diag}\times\urm(1)'~
\longrightarrow~\SU(r)_{\rm diag} \ ,
\eeq
and only $\SU(r)_{\rm diag}$ survives as a gauge symmetry. The
quadratic form (\ref{eq:AMA}) is given by
\beq 
\tr_{r\times r}\Bigl(\big(D\phi^0\big)^\dag\, D\phi^0\Bigr)=
\frac 1{8R^2}\,\left[2\delta_{ab}\,\left(\mbox{$\frac 1{\sqrt{2}}$}\,
A^a_L - A^a_R\right)\, \left(\mbox{$\frac 1{\sqrt{2}}$}\,
A^b_L - A^b_R\right) +\mbox{$\frac 3 r$}\, B^2 \right] \ .
\eeq
The gauge boson mass matrix is thus given by
\beq
\mbf M^2=\frac 1 {2R^2}\,\begin{pmatrix} \frac 1 2 ~{\bf 1}_r & -\frac
  1 {\sqrt{2}}~ {\bf 1}_r  &  0 \\ 
-\frac 1 {\sqrt{2}}~ {\bf 1}_r & {\bf 1}_r & 0 \\
0 & 0 & \frac 3 {2r}\, \\
\end{pmatrix} \ .
\label{M2r}\eeq
Diagonalising (\ref{M2r}) produces massive gauge bosons $B$ together
with
\beq
W^a:=\sqrt{\mbox{$\frac 1 3$}}~ A^a_L - \sqrt{\mbox{$\frac 23$}}~A^a_R
\eeq
with mass squared
\beq
\mu^2_B\=\frac 3 {4r\, R^2} \qquad \mbox{and} \qquad
\mu^2_W\=\frac 3{4 R^2} \ ,
\eeq
while the massless combinations corresponding to the unbroken symmetry
group $\SU(r)_{\rm diag}$ are
\beq
A^a:=\sqrt{\mbox{$\frac 23$}}~ A^a_L + \sqrt{\mbox{$\frac 1 3$}}~
A^a_R \ .
\eeq
The physical Higgs fields can be incorporated into an $r\times r$
hermitean matrix $h$ with
\beq
\phi=\frac 1{\sqrt{2}\,g \,R}~{\bf 1}_r + h \ ,
\eeq
and the Higgs boson mass read off from the term in the Higgs potential
(\ref{Higgspotchain}) quadratic in $h$ to get
\beq
\mu^2_h=\frac 6{R^2} \ .
\label{Higgsmassr}\eeq

\bigskip

\noindent
{\bf $\mbf{p_0=1 \ , \ p_1=2}$ . \ } 
This example exhibits $\urm(1)$ mixing. One has $p=5$ and the pattern
\beq
\SU(5)~\longrightarrow~\SU(2)\times\urm(1)~\longrightarrow~\urm(1)' \
.
\label{SU5break}\eeq
In this case the Higgs field $\phi$ is a two-component column vector
with vacuum expectation value
\beq 
\phi^0=\frac 1{\sqrt 2\, g\,R}\,\begin{pmatrix}  0 \\
  1  \end{pmatrix} \ .
\eeq
The Higgs boson mass is again given by (\ref{Higgsmassr}), but
now the gauge boson mass matrix obtained from (\ref{eq:Dphi}) and
(\ref{eq:AMA}) mixes $A_L^3$ and $B$ as
\beq
\mbf M^2=\frac 1 {8R^2}\,\begin{pmatrix} \, 1 & 0 & 0 & 0 \\ 0 & 1 & 0
  & 0 \\ 0 & 0 & 1 & -\sqrt 5 \\ 0 & 0 & -\sqrt5 & 5
\end{pmatrix} \ .
\eeq
This gives two $W$-bosons with mass squared
\beq
\mu_W^2=\frac 1 {8 R^2} \ ,
\eeq
a $Z$-boson
\beq
Z\=\mbox{$\frac 1 {\sqrt 6}$}\,  \big(A_L^3 - \sqrt 5~
B\,\big) \qquad  \mbox{with} \quad \mu_Z^2\=\frac 3
{4R^2} \ ,
\eeq
and a massless photon
\beq
A=\mbox{$\frac 1 {\sqrt 6}$}\,\big(\,\sqrt 5~
A_L^3 + B\,\big)\ .
\eeq
The Weinberg angle $\theta$ in this model is given by
\beq
\sin^2 \theta=\frac 5
6 \ . 
\eeq

\bigskip

\noindent
{\bf $\mbf{p_0=2 \ , \ p_1=1}$ . \ } 
Here $p=4$ and the symmetry breaking sequence (\ref{SU5break}) is
modified to
\beq
\SU(4)~\longrightarrow~\SU(2)\times\urm(1)~\longrightarrow~\urm(1)' \
.
\eeq
In this case one computes
\beq
\mu_W^2\=\frac 1 {4R^2} \ , \qquad \mu_Z^2\=\frac 3 {4R^2} \qquad
\mbox{and} \qquad \sin^2\theta \= \frac 2 3 \ .
\eeq
This example illustrates that, in contrast to the $\CP^1$ case, the
results depend on the ordering of the quiver gauge group ranks $p_n$.

\subsection{Fermion spectrum and Yukawa couplings}

Following the general analysis of \S\ref{Yukawa} and
\S\ref{YMHchains}, there are two fermion zero modes $\psi_{0,-2}$ and
$\widetilde\psi_{1,1}$ on $M$ determined by the twisting parameter
$\ct=-\frac 12$, for which the index is given by
\beq
\nu_{1,1}\=-1 \qquad \hbox{and} \qquad \nu_{0,-2}\=+1 \ .
\eeq
The positive chirality mode on $\CP^2$ is
\beq
\chi^+_{0,-2}=|\Omega\rangle \otimes \big|{\stackrel{\scriptstyle 0}
  {\scriptstyle 0}}\,,\,-2\big\rangle
\eeq
while the negative chirality mode, which is a doublet of the $\SU(2)$
gauge theory on $\CP^2$, is
\beq 
\chi^-_{1,1}=\frac 1{\sqrt 2}\,\Big(\sigma^{\bar 1}
|\Omega\rangle\otimes\big|{\stackrel{\scriptstyle 1}
{\scriptstyle 1}}\,,\,1\big\rangle+\sigma^{\bar 2}
|\Omega\rangle\otimes\big|{\stackrel{\scriptstyle 1}
{\scriptstyle -1}}\,,\,1\big\rangle\Big) \ .
\eeq
For example, taking $p_0=p_1=r$, we can choose the corresponding
$d$-dimensional spinor fields $\widetilde\psi_{1,1}$ and $\psi_{0,-2}$
to transform in the fundamental representation of
$\SU(r)\times\SU(r)$. After the rescalings (\ref{phiArescale}) and
(\ref{fermrescale}), the Yukawa couplings in (\ref{EDdimred}) for this
case take the form
\beq
2g\,\int_{\CP^2}\,\beta_{\rm vol}~ \Psi^\dag
 \begin{pmatrix} 0 &0 & \phi\,\gamma\otimes\sigma^{\bar 1}\,\\ 0 &0 &
   \phi\,\gamma\otimes\sigma^{\bar 2}\\
   \,\phi^\dag\,\gamma\otimes\sigma^1 &\phi^\dag\,\gamma\otimes\sigma^2
   & 0\end{pmatrix} 
\Psi= 2 g\, \bigl(\widetilde\psi_{1,1}^{\,\dag}\,\phi\,\gamma
\,\psi_{0,-2} + \psi_{0,-2}^\dag\, \phi^\dag\,\gamma
\,\widetilde\psi_{1,1}\bigr)
\eeq
where we have used (\ref{Psipsichi}). Expanding about the Higgs vacuum
(\ref{HiggsVEV10}), we find a mass term for the $d$-dimensional
fermions given by
\beq\label{FundamentalYukawa}
\frac {\sqrt{2}} R \,\Bigl(\psi_{1,1}^\dag \,\psi_{0,-2} +
\psi_{0,-2}^\dag \,\psi_{1,1}\Bigr) \ ,
\eeq
where $\psi_{1,1}=\gamma \,\widetilde \psi_{1,1}$. This agrees with
(\ref{muk0}) for $k=1$.

In addition to the zero modes there is an infinite tower of massive
modes. The full spectrum of the Dirac operator on $\CP^2$ can be
derived using the results of \S\ref{Yukawa}. For this, we require the
irreducible $\SU(2)\times\urm(1)$ representations that appear in the
tensor product of the gauge group representations $\underline{(1,1)}$
and $\underline{(0,-2)}$ on $\CP^2$ with the spinor representation
(\ref{SpinorRep}), which was shown in \S \ref{Yukawa} to decompose as
$[\,\underline{(1,0)}\,]~\oplus~[\,\underline{(0,3)}\,\oplus
\,\underline{(0,-3)}\,]$. Twisting with $c=-1$ from
(\ref{twistunique}), to give globally well-defined bundles, alters the 
gauge group representations as $\underline{(1,1)}\rightarrow
\underline{(1,0)}$ and $\underline{(0,-2)}\rightarrow
\underline{(0,-3)}$. Thus we require the eigenvalues, and their multiplicities,
of the Dirac operator for the representations
\beq 
\underline{(1,0)}\,\otimes\, \big(\,\big[\,\underline{(1,0)}\,\big]~
\oplus~\big[\,\underline{(0,3)}\,\oplus\, \underline{(0,-3)}\,\big]\,
\big)=\big[\,\underline{(2,0)}\,\oplus\,\underline{(0,0)}\,\big]~
\oplus~\big[\,\underline{(1,3)}\,\oplus\,\underline{ (1,-3)}\,\big]
\label{Decomp1}
\eeq
and
\beq 
\underline{(0,-3)}\,\otimes \,\big(\,\big[\,\underline{(1,0)}\,\big]~
\oplus~\big[\,\underline{(0,3)}\,\oplus\, \underline{(0,-3)}\,\big]\,
\big)=\big[\,\underline{(1,-3)}\,\big]~\oplus~\big[\,
\underline{(0,0)}\,\oplus\,\underline{(0,-6)}\,\big] \ .
\label{Decomp2}
\eeq

The eigenvalues and their multiplicities follow from the general
formulas (\ref{DiracEigenvalues1}) and (\ref{DiracEigenvalues2}) of
\S\ref{Yukawa}. The eight states on the right-hand side of
(\ref{Decomp1}), a triplet, two doublets and a singlet of $\SU(2)$,
give rise to eight infinite sequences of Dirac eigenspinors. All
eigenvalues occur in equal pairs with opposite sign so there are four
infinite sequences with positive eigenvalues, together with their
negative eigenvalue partners. The four states on the right-hand side
of (\ref{Decomp2}), a doublet and two singlets of $\su$, give rise to
four infinite sequences of Dirac eigenspinors with eigenvalues in
equal pairs and opposite signs yielding two infinite sequences with
positive eigenvalues, together with their negative eigenvalue
partners. Denoting the positive eigenvalues 
by $\frac{\lambda_N} R $, with degeneracies $d_N$, the two infinite
sequences arising from (\ref{Decomp2}) are given by
\bea
\lambda_N&=&\sqrt{(N+1)\,(N+3)} \ ,\qquad d_N\=(N+2)^3 \ ,
\nonumber \\[4pt]
\lambda_N&=&\sqrt{(N+2)\,(N+3)} \ , \qquad d_N\=\mbox{$\frac 1 2$}\,
(N+1)\,(N+4)\,(2N+5)
\label{FundamentalSpectrum}\eea
with $N=0,1,\ldots$. The spectrum arising from (\ref{Decomp1}) gives
two copies of (\ref{FundamentalSpectrum}), so the full spectrum
consists of three copies of (\ref{FundamentalSpectrum}) together with
their negative eigenvalue counterparts. The two zero modes can be
thought of as coming from two copies of the first sequence in
(\ref{FundamentalSpectrum}) with $N=-1$.

It can be interesting to also consider alternative values of the
twisting  parameter $c$, other than the choice $c=-1$ which induces
Yukawa couplings in the zero mode sector of the fermionic field theory
on $M$. In the present context $c=3$ gives three positive chirality
zero modes, $\nu_{1,1}=3$ while $\nu_{0,-2}=0$, and $c=-3$ gives three
negative chirality zero modes, $\nu_{1,1}=0$ while
$\nu_{0,-2}=-3$. These zero modes could manifest themselves as three
generations of fermions in the dimensionally reduced field theory.

\bigskip

\section{Dynamical symmetry breaking  from the adjoint
  representation\label{Adjoint}}

\noindent
In this section we examine symmetry breaking from the
eight-dimensional adjoint representation $\underline{C}^{1,1}$ of
$\sut$. This is the lowest representation which is qualitatively
distinct from the $\CP^1$ examples, in the sense that it involves a
full two-dimensional quiver lattice (\ref{Figquiver}) of equivariant
gauge fields. Again we will determine the physical particle spectrum
and masses in some explicit instances.

\subsection{Spontaneous symmetry breaking}

In the case $k=l=1$, the weight set is $\quiver_{1,1}=
\big\{(1,3)\,,\,(1,-3)\,,\,(2,0)\,,\,(0,0)\big\}$. The only non-zero
coefficients $\Lambda^\pm_{1,1}(n,m)$ in (\ref{lambdaklnm}) are
\beq 
\Lambda^+_{1,1}(1,-3)\=1 \ , \qquad
\Lambda^+_{1,1}(0,0)\=\sqrt{\mbox{$\frac 3 2$}}\ , \qquad 
\Lambda^-_{1,1}(1,-3)\=\sqrt 3 \qquad\hbox{and}\qquad
\Lambda^-_{1,1}(2,0)\=\sqrt{\mbox{$\frac 3 2$}} \ .
\eeq
Hence the only four matrix one-forms in (\ref{betanmdef}) are
\beq
\b^\pm_{1,-3} \ , \qquad \b^+_{0,0} \qquad\hbox{and}\qquad
\b^-_{2,0} \ ,
\eeq
and there are only four Higgs fields
\beq 
\phi^\pm_{1,-3} \ , \qquad \phi^+_{0,0} \qquad\hbox{and}\qquad
\phi^-_{2,0} \ .
\eeq
The apparent asymmetry here, in that the weight $(1,3)$ does not
appear while $(1,-3)$ does, is an artifact of the notation. The
symmetry between the representations is clear in the quiver lattice
\beq
\mbox{\includegraphics[width=1.3in]{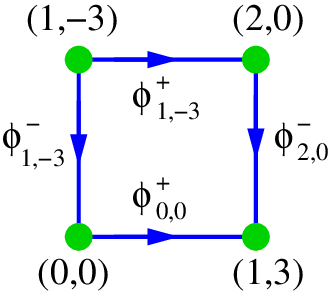}}
\label{Figadjoint}\eeq
that indicates which $\SU(2)\times\urm(1)$ representations are mapped
by the Higgs field morphisms. 

For illustrative purposes, we will again restrict to the case of equal
quiver gauge group ranks given by $p_{1,-3}=p_{1,3}=p_{0,0}=p_{2,0}=r$
with $p=8r$, which gives the gauge symmetry reduction pattern
$\SU(8r)\rightarrow\SU(r)^4\times\urm(1)^3$. 
In this case each Higgs field $\phi^\pm_{n,m}$ is a square $r\times r$
matrix and the Higgs potential in (\ref{SklYM}), after the rescalings
(\ref{phiArescale}), is
\bea V\big(\mphi^+\,,\,\mphi^-\big)&=&g^2~\tr_{r\times r}\bigg[\,
3\,\Big(\, \frac 3 {4g^2\,R^2}~ \mathbf{1}_r
  -\big(\phi^+_{0,0}\big)^\dag\, \phi^+_{0,0}\Big)^2 
+\frac 5 3\, \Big(\, \frac 3 {4g^2\,R^2}~ \mathbf{1}_r
  -\big(\phi^+_{1,-3}\big)^\dag\,\phi^+_{1,-3}\Big)^2 \nonumber \\ 
&& \qquad\qquad +\,3\,\Big(\, \frac 3 {4g^2\,R^2}~ \mathbf{1}_r
  -\big(\phi^-_{1,-3}\big)^\dag\, \phi^-_{1,-3}\Big)^2
+\frac 5 3 \,\Big(\, \frac 3 {4g^2\,R^2}~ \mathbf{1}_r
  -\big(\phi^-_{2,0}\big)^\dag\,\phi^-_{2,0}\Big)^2 \nonumber \\ 
&& \qquad\qquad +\,\Big| \phi^+_{1,-3} \,\big(\phi^-_{1,-3}\big)^\dag -
    \big(\phi^-_{2,0}\big)^\dag \,\phi^+_{0,0} \Big|^2
+ \Big| \phi^+_{0,0}\, \phi^-_{1,-3`} - \phi^-_{2,0}\, \phi^+_{1,-3}
\Big|^2 \, \bigg] \ . 
\label{AdjointPotential}\eea
The global minimum of (\ref{AdjointPotential}) is attained by setting
all four Higgs fields proportional to $\urm(r)$ matrices 
\beq 
\phi^\pm_{n,m}{}^0=\frac {\sqrt 3}{2 g \,R}~U^\pm_{n,m}\ ,
\label{adjointVeV}\eeq  
which is a special instance of (\ref{phinmpmexplgen}), together with
the constraint
\beq 
U^-_{2,0}\,U^+_{1,-3}= U^+_{0,0}\,U^-_{1,-3} \ .
\label{Uconstraint}\eeq
There are therefore only three independent unitary matrices
$U^\pm_{n,m}$, and we can use a $\urm(r)^3$ gauge transformation
to set any three of them equal to $\mathbf{1}_r$. The constraint
(\ref{Uconstraint}) then requires all four to be the identity and only
the diagonal subgroup $\SU(r)_{\rm diag}$ survives. The gauge symmetry
is thus broken dynamically as
\beq
\SU(8r)~\longrightarrow~\SU(r)^4\times\urm(1)^3~
\longrightarrow~\SU(r)_{\rm diag} \ ,
\eeq
with all four Higgs fields transforming in the same way under the
surviving diagonal subgroup as $\phi^\pm_{n,m} \rightarrow
g\,\phi^\pm_{n,m}\,g^\dag$ where $g\in\SU(r)_{\rm diag}$. Of the
initial $(4r^2-1)$ gauge bosons, $3r^2$ become massive and of the
original $8r^2$ degrees of freedom in the four complex Higgs fields,
$5r^2$ survive as physical Higgs fields.

We can parameterise the physical Higgs fields by choosing a gauge in
which three are given by hermitean matrices and one by a general
complex matrix, yielding $5r^2$ degrees of freedom as required.
To see that such a gauge exists, we first observe that any square
complex matrix has a unique polar decomposition into the product of a
unitary matrix with a hermitean matrix so that, without making any
gauge choice, we can always write
\beq 
\phi^\pm_{n,m}=V^\pm_{n,m}\,\Big(\,\frac
{\sqrt{3}}{2g\,R}~\mathbf{1}_r+ h^\pm_{n,m}\Big) 
\eeq
with $V^\pm_{n,m}$ unitary and $h^\pm_{n,m}$ hermitean. In this
parameterisation the vacuum state (\ref{adjointVeV}) corresponds to
$h^\pm_{n,m}=0$ and $V^\pm_{n,m}=U^\pm_{n,m}$ satisfying
(\ref{Uconstraint}). By using an $\SU(r)^4\times\urm(1)^3$ gauge
transformation we can set any three of the $\urm(r)$-valued fields
$V^\pm_{n,m}$ to the identity, but not all four. Let us choose a gauge
in which $V^\pm_{1,-3}=V^-_{2,0}=\mathbf{1}_r$. In this gauge, the
Higgs fields
\beq 
\phi^\pm_{1,-3}\=\frac {\sqrt{3}}{2g\, R}~\mathbf{1}_r+ 
h^\pm_{1,-3} \qquad \mbox{and} \qquad
\phi^-_{2,0}\=\frac {\sqrt{3}}{2g\, R}~\mathbf{1}_r+ 
h^-_{2,0}
\eeq
are hermitean while
\beq
\phi^+_{0,0}= V^+_{0,0}\,\Big(\,\frac {\sqrt{3}}{2g\, R}~
\mathbf{1}_r+ h^+_{0,0}\Big)
\label{Phi00Polar}
\eeq
is a general complex matrix. Although $V^+_{0,0}$ is an arbitrary
unitary field in general, the vacuum condition (\ref{Uconstraint}) in
this gauge requires $U^+_{0,0}=\mathbf{1}_r$ so let us paramaterise
$\phi^+_{0,0}$ differently. Instead of (\ref{Phi00Polar}), it will be
more convenient to use the decomposition 
\beq
\phi^+_{0,0}=\frac {\sqrt{3}}{2g\, R}~\mathbf{1}_r+ H^+_{0,0}+
\im \widetilde H^+_{0,0}
\eeq
with $H^+_{0,0}$ and $\widetilde H^+_{0,0}$ hermitean. In this gauge
the $5r^2$ physical degrees of freedom in the Higgs fields are
represented by the five hermitean matrices $h^\pm_{1,-3}$,
$h^-_{2,0}$, $H^+_{0,0}$ and $\widetilde H^+_{0,0}$, and the remaining
$\SU(r)_{\rm diag}$ gauge degree of freedom is implemented by
$(h_{n,m}^\pm,H^+_{0,0},\widetilde H^+_{0,0})\rightarrow g\,
(h_{n,m}^\pm,H^+_{0,0},\widetilde H^+_{0,0}) \,g^\dag $ with
$g\in\SU(r)_{\rm diag}$.

The Higgs boson masses can be found by extracting the quadratic part
of the potential (\ref{AdjointPotential}) when expanded around the
minimum. The mass matrix $\mbf M_h$ works out to be given by
\beq
\mbf M_h^2=\frac 1 {2R^2}\,\begin{pmatrix}
\, 3 & 0 & 0 & 0 & 0 \\ 
0 & 21 & -3 & 0 & 0 \\
0& -3 & 13 & 0 & 0 \\ 
0& 0 & 0 & 21 & -3  \\
0 & 0 & 0 & -3 & 13 \\
\end{pmatrix}\otimes\Idd_r \ ,
\eeq
where the rows and columns are labelled by the sequence of Higgs
fields $\big\{\widetilde H^+_{0,0},H^+_{0,0},
h^+_{1,-3},h^-_{1,-3},h^-_{2,0}\big\}$. 
There are two doubly degenerate eigenvalues
\beq
\mu^2_{h^\pm}=\frac {11}{R^2} 
\eeq
corresponding to the linear combinations
\beq
h^+\=\mbox{$\frac 1 {\sqrt{10}}$}\, \bigl( h^+_{1,-3} - 3
H^+_{0,0}\bigr) \qquad \mbox{and} \qquad
h^-\=\mbox{$\frac 1 {\sqrt{10}}$}\, \bigl( h^-_{2,0} - 3
h^-_{1,-3}\bigr) \ , 
\eeq
and
\beq
\mu^2_{h^{\prime\,\pm}}=\frac{6}{R^2}
\eeq
associated with the orthogonal combinations
\beq
h^{\prime\,+}\=\mbox{$\frac 1 {\sqrt{10}}$}\, \bigl( 3h^+_{1,-3} +
H^+_{0,0}\bigr) \qquad \mbox{and} \qquad
h^{\prime\,-}\=\mbox{$\frac 1 {\sqrt{10}}$}\, \bigl( 3h^-_{2,0} +
h^-_{1,-3}\bigr) \ .
\eeq
The lightest Higgs field is $\widetilde H^+_{0,0}$ with
\beq
\mu^2_{\widetilde H^+_{0,0}}=\frac {3}{2R^2} \ .
\eeq

The gauge boson masses are determined from the bicovariant derivative
terms in (\ref{SklYM}), after the rescalings (\ref{phiArescale}) and
setting $\phi^\pm_{n,m}$ equal to their vacuum expectation values.
Again writing the gauge potentials $A^{n,m}=\frac \im 2\,A^a_{n,m}\,
\lambda_a+\frac \im 2\,A^0_{n,m}\,\sqrt{2/ r}~\mathbf{1}_r$  in terms
of Gell-Mann matrices $\lambda_a$ for $\SU(r)$ and the identity
matrix, using (\ref{curvoffdiagCklrep}) one finds
\bea
D\phi^\pm_{n,m}&=&\diff\phi^\pm_{n,m}+\frac {\im g} 2 \,
\bigg(\,\frac{A_{n\pm 1,m+3}^a} {\sqrt {n+1\pm
    1}}\,\lambda_a\,\phi^\pm_{n,m} 
- \frac {A_{n,m}^a} {\sqrt{n+1}}\,\phi^\pm_{n,m}\,\lambda_a
\nonumber\\ && \qquad\qquad\qquad +\,\sqrt{\frac 2 r}~
\Big(\, \frac   {A_{n\pm 1,m+3}^0} {\sqrt {n+1\pm 1}} 
- \frac {A_{n,m}^0} {\sqrt{n+1}} \, \Big)\, \phi^\pm_{n,m}\bigg) \ .
\label{Dphinmadjoint}\eea
By defining the normalised $\urm(1)$ fields
\beq
B^\pm_{n,m}:=\frac 1 {\sqrt{2n+2\pm1}}\,\Big(\sqrt{n+1} \;A_{n\pm
  1,m+3}^0 -\sqrt{n+1\pm1}\;A_{n,m}^0\Big) \ , 
\label{Bpmnm}\eeq
we can rewrite (\ref{Dphinmadjoint}) as
\bea
D\phi^\pm_{n,m}&=&\diff\phi^\pm_{n,m}+\frac {\im g} 2\,
\bigg(\,\frac{A_{n\pm 1,m+3}^a} {\sqrt {n+1\pm
    1}}\,\lambda_a\,\phi^\pm_{n,m} 
- \frac {A_{n,m}^a} {\sqrt{n+1}}\,\phi^\pm_{n,m}\,\lambda_a
\nonumber\\ && \qquad\qquad\qquad +\,
\sqrt{\frac {2(2n+2\pm 1)} {r\,(n+1\pm 1)\,(n+1)}}~
B^\pm_{n,m}\, \phi^\pm_{n,m}\bigg) \ .
\eea
Not all four fields (\ref{Bpmnm}) are independent of course, as there
are only three $\urm(1)$ degrees of freedom, and indeed one has
\beq 
B^-_{1,-3}=-B^+_{0,0} \ .
\eeq

Now using (\ref{adjointVeV}) gives the quadratic form
\bea
\mbox{$\frac 1 2$}\,{\mbf A}^\top\,\mbf M^2\,{\mbf A}&=&\tr_{r\times
  r} \Bigl(\bigl(D\phi^{+}_{1,-3}{}^0\bigr)^\dag\,D\phi^{+}_{1,-3}{}^0
+  \bigl(D\phi^{+}_{0,0}{}^0\bigr)^\dag\,D\phi^{+}_{0,0}{}^0
\nonumber\\ && \qquad\qquad +\, 
\bigl(D\phi^{-}_{1,-3}{}^0\bigr)^\dag\,D\phi^{-}_{1,-3}{}^0 + 
\bigl(D\phi^{-}_{2,0}{}^0\bigr)^\dag\,D\phi^{-}_{2,0}{}^0\Bigr)
\eea
with the gauge boson mass matrix given by
\beq
\mbf M^2=\frac 3{4R^2}\,
\begin{pmatrix}
\mathbf 1_r & \mathbf 0_r & -\sqrt{\frac 1 2}~ \mathbf 1_r &
-\sqrt{\frac 1 6}~ \mathbf 1_r & 0 & 0 & 0 \\ 
\mathbf 0_r & \mathbf 1_r & -\sqrt{\frac 1 2}~ \mathbf 1_r &
-\sqrt{\frac 1 6}~ \mathbf 1_r & 0 & 0 & 0 \\ 
-\sqrt{\frac 1 2}~ \mathbf 1_r & -\sqrt{\frac 1 2}~ \mathbf 1_r &
2~\mathbf 1_r & \mathbf 0_r  & 0 & 0 & 0 \\ 
-\sqrt{\frac 1 6}~ \mathbf 1_r & -\sqrt{\frac 1 6}~ \mathbf 1_r &
\mathbf 0_r & \frac 2 3 ~\mathbf 1_r & 0 & 0 & 0\\ 
0 & 0 & 0 & 0 & 3 & 0 & 0 \\
0 & 0 & 0 & 0 & 0 & \frac 5 6 & 0 \\
0 & 0 & 0 & 0 & 0 & 0 & \frac 5 6 \,
\end{pmatrix} \ ,
\eeq
where the rows and columns of the mass matrix are ordered according to
the sequence of gauge potentials $\big\{{A}^{1,3}, {A}^{1,-3},
{A}^{0,0}, 
 {A}^{2,0},B^-_{1,-3},B^+_{1,-3},B^-_{2,0}\big\}$.
The eigenvalues of the upper left $4\times4$ block matrix are
\beq
0 \ , \qquad \frac 2{R^2} \qquad \mbox{and} \qquad \frac 3 {4R^2}
\quad \mbox{(twice)} \ .
\eeq
The linear combination 
\beq
A_a:=\mbox{$\frac 1 2$}\,\Bigl(A^{1,3}_a
+A^{1,-3}_a+\sqrt{\mbox{$\frac 1 2$}}~A^{0,0}_a +\sqrt{\mbox{$\frac 3
    2$}}~A^{2,0}_a\Bigl)  
\label{SUrdiag}\eeq 
is massless, while the gauge boson
\beq 
\mbox{$\frac 1 2\,\sqrt{\frac 1 {10}}$}\,\Bigl(-\sqrt{6}\,
\big(A^{1,3}_a +A^{1,-3}_a\big)+3\,\sqrt{3}~A^{0,0}_a +A^{2,0}_a\Bigl)
\eeq 
has mass squared $\frac 2 {R^2}$. The two linear combinations with
mass squared $\frac 3 {4R^2}$ are
\beq
\sqrt{\mbox{$\frac 1 2$}}\,\Bigl(A^{1,3}_a - A^{1,-3}_a\Bigr) 
\qquad \mbox{and} \qquad
\sqrt{\mbox{$\frac 1 {10}$}}\,\Bigl(A^{1,3}_a +A^{1,-3}_a+\sqrt{2}~
A^{0,0}_a -\sqrt{6}~A^{2,0}_a\Bigl) \ .
\eeq 
In addition, the three $\urm(1)$ gauge bosons acquire masses given by
\beq 
\mu^2_{B^-_{1,-3}}\=\frac 9 {4R^2} \qquad \hbox{and} \qquad
\mu^2_{B^+_{1,-3}}\=\mu^2_{B^-_{2,0}}\=\frac 5 {8R^2} \ .
\eeq
It seems remarkable that the mass squared for all Higgs bosons and
gauge bosons evaluate to rational multiples of $\frac 1 {R^2}$. 

\subsection{Fermion spectrum and Yukawa couplings}

Following the analysis of \S\ref{Yukawa}, with twisting parameter
$c=-3$ there is a positive chirality zero mode associated with the
$\su$ singlet
\beq
\chi^+_{0,0} =|\Omega\rangle\otimes\bigl|{\stackrel{\scriptstyle 0} 
  {\scriptstyle 0}}\,,\,0\big\rangle \ ,
\eeq
and a negative chirality mode associated with one of the $\su$
doublets
\beq \chi^-_{1,3} =\frac1{\sqrt2}\,\Big(\sigma^{\bar 1}
|\Omega\rangle\otimes\big|{\stackrel{\scriptstyle 1}
{\scriptstyle 1}}\,,\,3\big\rangle+\sigma^{\bar 2}
|\Omega\rangle\otimes\big|{\stackrel{\scriptstyle 1}
{\scriptstyle -1}}\,,\,3\big\rangle\Big) \ .
\eeq
If a $(d+4)$-dimensional spinor field $\Psi$ transforms in the fundamental 
representation of $\SU(8r)$, then the $d$-dimensional spinors
$\psi_{0,0}$ and $\psi_{1,3}=\gamma\widetilde\psi_{1,3}$, associated
with $\chi^+_{0,0}$ and $\chi^-_{1,3}$ respectively, transform under
fundamental representations of the different $\SU(r)$ gauge groups
with connections $A^{0,0}$ and $A^{1,3}$. When the quiver gauge
symmetry is broken, they both transform under the fundamental
representation of the remaining unbroken $\SU(r)_{\rm diag}$
combination, with respective charges $\frac g {2\,\sqrt{2}}$ and
$\frac g2$ according to~(\ref{SUrdiag}). From
(\ref{mukl}) it follows that the Yukawa couplings give masses
$\mu_{1,1}$ to these fermions with
\beq 
\mu_{1,1}^2=\frac 9 {2R^2} \ .
\eeq

By (\ref{nunm}), the index associated with the weight $(n,m)=(2,0)$ is
zero, but the index for $(n,m)=(1,-3)$ is $\nu_{1,-3}=8$. Thus unlike
the fundamental representation breaking, the adjoint representation
breaking models contain massless chiral fermions. We can expect the
same to be true for all representations $\underline{C}^{k,l}$ with
$k+l>1$ when $l>0$, and with $k>2$ when $l=0$ (see (\ref{SDk0})).

Again there is an infinite tower of massive Dirac eigenspinors.   
Twisting with $c=-3$ alters the weights in $\quiver_{1,1}$ as
\beq
(1,3)~\longrightarrow~(1,0) \ , \qquad
(1,-3)~\longrightarrow~(1,-6) \ , \qquad
(2,0)~\longrightarrow~(2,-3) \qquad \mbox{and} \qquad
(0,0)~\longrightarrow~(0,-3)
\eeq
and the corresponding $H$-modules are then tensored with the spinor
representation, as in (\ref{SpinBundleDecomposition}), to yield 12
irreducible holonomy group representations given by the decompositions
\bea \label{AdjointSpinorDecomposition}
\underline{(1,0)}\,\otimes
\,\big(\,\big[\,\underline{(1,0)}\,\big]~\oplus~
\big[\,\underline{(0,3)}\,\oplus\,\underline{(0,-3)}\,\big]\,\big)&=& 
\big[\,\underline{(2,0)}\, \oplus\, \underline{(0,0)}\,\big]~ \oplus~
\big[\,\underline{(1,3)}\, \oplus\, \underline{(1,-3)}\,\big]  \ ,
\nonumber \\[4pt] 
\underline{(1,-6)}\,\otimes\,
\big(\,\big[\,\underline{(1,0)}\,\big]~\oplus
~\big[\,\underline{(0,3)}\,\oplus\,\underline{(0,-3)}\,\big]\,\big)&=& 
\big[\,\underline{(2,-6)}\, \oplus\, \underline{(0,-6)}\,\big]~
\oplus~ \big[\,\underline{(1,-3)}\, \oplus \,\underline{(1,-9)}\,\big] 
\ , \nonumber \\[4pt]
\underline{(2,-3)}\,\otimes\,
\big(\,\big[\,\underline{(1,0)}\,\big]~\oplus~
\big[\,\underline{(0,3)}\,\oplus\,\underline{(0,-3)}\,\big]\,\big)&=& 
\big[\,\underline{(3,-3)}\, \oplus\, \underline{(1,-3)}\,\big] ~\oplus 
~\big[\,\underline{(2,0)}\, \oplus\, \underline{(2,-6)}\,\big]  \ ,
\nonumber \\[4pt] 
\underline{(0,-3)}\,\otimes\,
\big(\,\big[\,\underline{(1,0)}\,\big]~\oplus~
\big[\,\underline{(0,3)}\,\oplus\,\underline{(0,-3)}\,\big]\,\big)&=& 
\big[\,\underline{(1,-3)}\,\big] ~\oplus~ \big[\,\underline{(0,0)}\,
\oplus\, \underline{(0,-6)}\,\big] \ . 
\eea
Since the total number of states in $\underline{C}^{1,1}$ is eight and
the spinor representation (\ref{SpinorRep}) is four-dimensional, there
are 32 infinite sequences corresponding to the 32 states on the
right-hand side of (\ref{AdjointSpinorDecomposition}). These consist
of 16 sequences of positive eigenvalues and their negative eigenvalue
partners. The 16 infinite sequences of positive eigenvalues
$\frac{\lambda_N}R$, together with their degeneracies $d_N$, arising
from the representations on the right-hand side of
(\ref{AdjointSpinorDecomposition}) can be calculated as before using
(\ref{DiracEigenvalues1}) and (\ref{DiracEigenvalues2}). They are
given by
\bea
\lambda_N&=&\sqrt{(N+1)\,(N+3)-2} \ ,\qquad \ d_N \= (N+2)^3 \ ,
\nonumber\\[4pt] 
\lambda_N&=&\sqrt{(N+1)\,(N+3)} \ ,\qquad \qquad d_N \= (N+2)^3\hskip
105pt \ (\times 3) \ , \nonumber\\[4pt]  
\lambda_N&=&\sqrt{(N+2)\,(N+3)-3} \ ,\qquad \  d_N\=\mbox{$\frac 1
  2$}\, (N+1)\,(N+4)(2N+5)  \ , \nonumber\\[4pt] 
\lambda_N&=&\sqrt{(N+2)\,(N+3)-2} \ ,\qquad  \ d_N\=\mbox{$\frac 1
  2$}\, (N+1)\,(N+4)\,(2N+5) \qquad (\times 2) \ , \nonumber\\[4pt] 
\lambda_N&=&\sqrt{(N+2)\,(N+3)} \ ,\qquad  \qquad d_N\=\mbox{$\frac 1
  2$}\,(N+1)\,(N+4)\,(2N+5) \qquad (\times 3)
\label{ZeroModeSeries}\eea
and
\bea
\lambda_N&=&\sqrt{(N+1)\,(N+5)} \ ,\qquad \qquad d_N \=(N+3)^3   \ ,
\nonumber\\[4pt] 
\lambda_N&=&\sqrt{(N+1)\,(N+5)+1} \ ,\qquad  \ d_N \=(N+3)^3  \ ,
\nonumber\\[4pt] 
\lambda_N&=&\sqrt{(N+2)\,(N+5)-1} \ ,\qquad \ d_N \=
\mbox{$\frac 1 2$}\, (N+2)\,(N+5)\,(2N+7) \ , \nonumber\\[4pt] 
\lambda_N&=&\sqrt{(N+2)\,(N+5)} \ ,\qquad \qquad d_N
\=\mbox{$\frac 1 2$}\, (N+2)\,(N+5)\,(2N+7)  \ , \nonumber\\[4pt] 
\lambda_N&=&\sqrt{(N+4)^2-1} \ ,\qquad \qquad \quad \ d_N \=
(N+1)\,(N+4)\,(N+7) \ , \nonumber\\[4pt] 
\lambda_N&=&N+4 \ , \hskip 92 pt \ \ \ d_N \= (N+1)\,(N+4)\,(N+7)
\label{Octet}\eea
with $N$ a non-negative integer. The two singlet zero modes are given
by setting $N=-1$ in two of the three sequences in the second line of
(\ref{ZeroModeSeries}), while the octet of zero modes is gotten by
taking $N=-1$ in the first sequence of (\ref{Octet}). 

\bigskip

\section{Conclusions\label{Conclusions}}

\noindent
We have examined in some detail the $\sut$-equivariant dimensional
reduction of pure massless Yang-Mills-Dirac theory over the coset
space $\C P^2$, including a systematic incorporation of monopole and
instanton backgrounds on $\CP^2$. The topologically non-trivial
internal fluxes induce a Higgs potential as well as Yukawa couplings
between the reduced fermion fields and the Higgs fields, with the
standard form of dynamical symmetry breaking. For the class of models
in which all Higgs fields are square matrices of the same dimension
$r$, the minima of the Higgs potential have a geometrical
interpretation in terms of gauge fields on the corresponding quiver
lattice. As a $\urm(r)$ lattice gauge theory configuration, the
non-abelian flux on the quiver lattice must vanish for the Higgs
vacuum to be realised. Explicit examples have been presented with
symmetry breaking hierarchies generated from both the fundamental and
adjoint representations of $\sut$.

For the fundamental representation models, the symmetry hierarchies
\bea
\SU(3r)&\longrightarrow&\SU(r)\times\SU(r)\times\urm(1)~
\longrightarrow~\SU(r) \ , \nonumber \\[4pt]
\SU(5)&\longrightarrow&\SU(2)\times\urm(1)~
\longrightarrow~\urm(1) \ , \\[4pt]
\SU(4)&\longrightarrow&\SU(2)\times\urm(1)~
\longrightarrow~\urm(1) \nonumber
\eea
have been analysed in detail, where the first symmetry breaking is
explicit, dictated by the equivariant dimensional reduction ansatz,
and the second one is dynamical. Gauge boson and Higgs masses have
been calculated in all three cases, and all are inversely proportional
the length scale set by the metric on $\CP^2$. The complete fermion
spectrum has been presented, including both chiral zero modes of the
Dirac operator and massive Dirac eigenmodes. There are two zero modes,
one of positive chirality $\chi^+_{0,-2}$ and one of negative
chirality $\chi^-_{1,1}$, which acquire masses via their Yukawa
couplings (\ref{FundamentalYukawa}), with left and right chiralities
of a single massive fermion carrying different $\SU(2)\times\urm(1)$
quantum numbers. This is analogous to the way that leptons and quarks
acquire masses in the standard model, with the left-handed and
right-handed electrons carrying different quantum numbers. The induced
zero mode masses are of the same order as the mass scale of the
infinite fermionic tower arising from the non-zero eigenvalues
(\ref{FundamentalSpectrum}). The infinite tower may be truncated to
finitely many degrees of freedom by replacing the coset space $\CP^2$
with a fuzzy projective plane $\CP^2_F$. However, while fuzzy versions
of the line bundle zero modes $\chi^+_{0,-2}$ are
known~\cite{DiraconFuzzyCP2}, there is as yet no explicit fuzzy
construction of zero modes on instanton bundles, though one certainly
exists. Models with realistic numbers of fermion generations can be obtained
by changing the spin$^c$ twisting parameter of~\S\ref{Yukawa}.

For the adjoint representation models, we examined the symmetry
breaking hierarchy 
\beq 
\SU(8r)~\longrightarrow~\SU(r)^4\times\urm(1)^3
~\longrightarrow~\SU(r) 
\eeq
in detail, calculating the gauge boson and physical Higgs masses
explicitly. Again chiral zero modes $\chi^+_{0,0}$ and $\chi^-_{1,3}$
of the Dirac operator exist for which masses are generated
by the Yukawa couplings. In this case, however, there is also an octet
of positive chirality zero modes which remains exactly massless. The
infinite tower of massive fermions obtained here is much more complicated than
that in the case of reductions over $\CP^1$, primarily because each state of a pertinent
irreducible representation of the isospin subgroup of the holonomy
group of $\CP^2$ generates an infinite tower of its own. For the
$\urm(1)$ holonomy group of $\CP^1$ all irreducible representations
are one-dimensional and there is only a single infinite tower for each
irreducible representation, while for $\CP^2$ any given irreducible
representation of $\SU(2)$ produces a family of infinite towers with
the number of members growing as the dimension of the
representation. Again these towers could be truncated by restricting
to a finite number of degrees of freedom using a fuzzy regularisation on $\CP^2_F$.

Many of the qualititative features we have unveiled regarding the
vacuum structure of field theories obtained via equivariant
dimensional reduction can be expected to hold over generic homogeneous
internal spaces $G/H$. The general structure of the induced quiver
gauge theories is described in~\cite{ACGP1,LPS1}. The quiver diagram
can be regarded as a lattice of dimension given by the rank of the holonomy
group $H$ of the coset, and it comes with relations which equate the
various distinct paths between any pair of vertices of the plaquettes
of the quiver lattice. These relations will arise dynamically as
conditions for the Higgs vacua. Thus, for instance, our lattice gauge
theoretic interpretation of the Higgs minima in terms of flat
connections will be a generic feature of any
coset space $G/H$ for which ${\rm rank}(H)\geq2$. With this in mind,
it would be interesting to extend our techniques to the equivariant
dimensional reductions of ten-dimensional ${\cal N}=1$ supersymmetric
${\rm E}_8$ gauge theories over six-dimensional coset
spaces~\cite{KZ1,Lopes1} and of superstring theories on nearly
K\"ahler backgrounds~\cite{Lopes2}. The most interesting class of such
reductions involve non-symmetric (and nearly K\"ahler) six-dimensional
coset spaces, with the vacua controlled by sets of torsion
fluxes. Presumably these internal fluxes could be systematically
incorporated, along with other topologically non-trivial background
fields of the coset space, in a manner analogous to the treatment of
this paper. More generally, it would be interesting to find internal
coset spaces for which the equivariant dimensional reduction leads to
a physical particle spectrum which is in more precise quantitative
agreement with that of the standard model.

\bigskip

\section*{Acknowledgments}

\noindent
B.P.D. wishes to thank the Dublin Institute of Advanced Studies for
financial support, and also the Perimeter Institute for Theoretical
Physics, where this work was completed, for their hospitality and
support. The work of R.J.S. was supported in part by the
EU-RTN Network Grant MRTN-CT-2004-005104 .

\bigskip

\appendix

\section{Bundles on $\CP^2$\label{ChernCharacter}}

\noindent
Consider the vector bundle $\Qcal$ over $\CP^2$ of rank two which is
inverse to the line bundle $\Lcal_{-1}$ with first Chern number $-1$,
in the K-theoretic sense
\beq 
\Qcal\oplus \Lcal_{-1}=I^3
\label{Qdef}
\eeq
where $I^3$ is the trivial bundle of rank three over $\CP^2$. The
bundle $\Qcal$ is called a quotient bundle~\cite{Bott+Tu}, and it has
structure group $\urm(2)$. Canonical connections on $\Lcal$ and
$\Qcal$ were given in (\ref{CP2moncurvexpl}) and (\ref{Fu2reduce}),
respectively, and indeed the construction of the flat connection $A_0$
in \S\ref{canconnCP2} was based on the decomposition
(\ref{Qdef}), see~\cite{LPS3}.

The Chern character of any bundle $\Vcal\to\CP^2$ of rank $r$
can be expanded as~\cite{Bott+Tu}
\beq 
\ch(\Vcal)=r + c_1(\Vcal) + \bigl(\mbox{${\frac 12}$}\,c_1(\Vcal)\wedge
c_1(\Vcal)-c_2(\Vcal)\bigr) \ ,
\eeq
where $c_1(\Vcal)$ and $c_2(\Vcal)$ are the first and second Chern
characteristic classes of $\Vcal$ with the integer
$C_2(\Vcal)=\int_{\CP^2} \,c_2(\Vcal)$ the second Chern number. The
rank two bundle $\Qcal$ carries $\urm(1)$ (magnetic monopole)
charge. Under the embedding
$\SU(2)\times\urm(1)\hookrightarrow\SU(3)$, the fundamental
representation of $\SU(3)$ decomposes as in
(\ref{fundrepdecomp}). This is the representation content of
(\ref{Qdef}). The line bundle $\Lcal_{-1}$ has first Chern number $-1$
and its fibres transform as the $H$-module
$\underline{(n,m)}=\underline{(0,-2)}$. The $\urm(1)$ quantum number
$m$ is thus twice the Chern number of the associated line bundle and
we shall call $\frac m2$ the monopole charge. The fibres of the
quotient bundle $\Qcal$ transform as the $H$-module
$\underline{(n,m)}=\underline{(1,1)}$. This implies that $\Qcal$ has
monopole charge $\frac 12$ but first Chern number $+1$, since it is of
rank two and the first Chern number involves a trace, so it is equal
to twice the monopole charge.

Chern characters are additive under Whitney sums of bundles, so since
$\Qcal\oplus\Lcal_{-1}$ is trivial we have
\beq
\ch(\Qcal\oplus \Lcal_{-1})\=\ch(\Qcal)+\ch(\Lcal_{-1})\=3
\label{QtoL}\eeq
giving $\ch(\Qcal)=3 -\ch(\Lcal_{-1})$. The Chern character is also
multiplicative with respect to tensor products of bundles, so using
(\ref{QtoL}) we have 
\beq
\ch(\Qcal\otimes \Lcal_\mt)\=\ch(\Qcal)\wedge\ch(\Lcal_\mt)
\=3\,\ch(\Lcal_\mt) - \ch(\Lcal_{\mt-1})
\label{chQLprod}\eeq
for any power $\mt$. In particular, for $\mt=-\frac 12$ we get the
instanton bundle $\Ical=\Qcal\otimes\Lcal_{-1/2}$ with
\begin{equation}
\ch(\Ical)=3\,\ch(\Lcal_{-1/2}) - \ch(\Lcal_{-3/2}) \ .
\end{equation}
The Chern character of the monopole line bundle $\Lcal$ is
$\ch(\Lcal)=\exp\xi$, where $\xi={\frac \im{2\pi}}\, f_\uoL$ with
$\int_{\CP^2}\,\xi\wedge\xi=-1$, so 
\beq
\ch(\Lcal)\=1+ \xi +\mbox{$\frac 12$}\,\xi\wedge\xi
\qquad \mbox{and} \qquad \int_{\CP^2}\,\ch(\Lcal)\=
-\frac 12 \ . 
\eeq
Similarly, one has
\beq 
\ch(\Ical)\=3\big(1-\mbox{$\frac 12\,\xi + \frac 18\,\xi
\wedge\xi\big) - \big(1-\frac 32\,\xi + \frac 98\,\xi\wedge
\xi\big)\=2-\frac 34\,\xi\wedge\xi$} 
\quad \mbox{and} \quad
\int_{\CP^2}\,\ch(\Ical)\=\frac 34 \ ,
\eeq
and hence the second Chern number of $\Ical$ is $\frac 34$, implying
that $\Ical$ does not exist globally~\cite{BDCP2}. Nevertheless, it
plays a crucial role in the index theorem described in
\S\ref{ASItheorem}.

We now have enough information to calculate the Chern characteristic
classes of the rank~$n+1$ instanton bundle $\Ical_n$. 
The relevant component of $\ch(\Ical)$ for evaluating the
integral over $\CP^2$ involves the square of the curvature two-form,
so an explicit evaluation requires taking the
trace of the second order Casimir operator in the two-dimensional
vector representation of $\SU(2)$. The Casimir operator is ${\sf
  C}_2(2)=\frac 34~{\bf 1}_2$, and taking the trace gives 
a factor of $2$, so
\beq 
\int_{\CP^2}\,\ch(\Ical)=\frac 12\,{\sf C}_2(2)\,\Tr({\bf 1}_2) \ .
\label{ChernFundamentalI}
\eeq
The bundle
\beq
\Ical_n:={\rm Sym}^{\otimes n}(\Ical)
\label{SymnIcal}\eeq
is the rank $(n+1)$ bundle given by the $n$-th symmetric tensor
product of $\Ical$. As such, its second Chern number differs from
(\ref{ChernFundamentalI}) in two ways. Firstly, the dimension 
of the fibre is $\Tr(\Idd_{n+1})$ and, secondly, the second order
Casimir operator is ${\sf C}_2(n+1)=\frac n2 \,(\frac n2
+1)~\Idd_{n+1}$. From this we deduce that the second Chern number of
$\Ical_n$ is  
\beq 
C_2(\Ical_n)\=-\int_{\CP^2}\,\ch(\Ical_n) \= -\frac 12\, 
\frac{n\,(n+2)}4 \,(n+1)\= -\frac 12\, I\,(I+1)\, (2I+1) \ .
\eeq
For spinor representations ($n=2I$ with $I\in\Z+\frac12$) this is
always fractional, while for vector representations ($n=2I$ with
$I\in\Z$) it is an integer corresponding to the dimension of the
irreducible $\sut$-representation $\underline C^{I,I-1}$.

\bigskip

\section{Matrix one-form products on $\CP^2$\label{betamatrixprods}}

\noindent
We record here the explicit matrix products which are used for
calculations in the quiver gauge theory of
\S\ref{Eqgaugeth}. Using (\ref{betanmdef}) the matrix one-form
products appearing in (\ref{curvdiagCklrep}) are given by
\bea
\betab^\pm_{n,m}{}^\dag\wedge\betab_{n,m}^\pm &=&
\frac{\Lambda_{k,l}^\pm(n,m)^2}{2(n+1)}~\Xi^\pm_+(n,m;\betab\,) \ ,
\label{betaprodCkl1}\\[4pt]
\betab^\pm_{n\mp1,m-3}\wedge\betab^\pm_{n\mp1,m-3}{}^\dag 
&=& -\,\frac{\Lambda_{k,l}^\pm(n\mp1,m-3)^2}{2(n+1\mp1)}~
\Xi^\pm_-(n,m;\betab\,) \ ,
\label{betaprodCkl2}\eea
where
\bea
\Xi^\pm_\varepsilon(n,m;\betab\,)&=&
\sum_{q\in\rel_n}\,\Big[\big((n\pm q+1\pm\varepsilon)~
\beta^1\wedge\betab^1
+(n\mp q+1\pm\varepsilon)~\beta^2\wedge\betab^2\big)~
\big|\noverq\,,\,m\big\rangle\big\langle\noverq\,,\,
m\big| \nonumber\\ &&  \hspace{2cm} +\,
\sqrt{(n+1)^2-(q+1)^2}~
\beta^1\wedge\betab^2~\big|\noverq\,,\,m\big\rangle
\big\langle{\stackrel{\scriptstyle n}{\scriptstyle q+2}}\,,\,m
\big|\nonumber\\ &&  \hspace{2cm} +\,
\sqrt{(n+1)^2-(q-1)^2}~
\beta^2\wedge\betab^1~\big|\noverq\,,\,m\big\rangle
\big\langle{\stackrel{\scriptstyle n}{\scriptstyle q-2}}\,,\,m
\big|\,\Big]
\label{Xinmbetab}\eea
with $\varepsilon=\pm\,1$ and $\Lambda_{k,l}^\pm(n,m):=0$ for
$n\leq0$. In (\ref{curvholrelsCklrep}) we encounter the matrix
one-form products
\beq
\betab^+_{n,m}\wedge\betab^-_{n+1,m-3}=\frac{
\Lambda_{k,l}^+(n,m)\,\Lambda_{k,l}^-(n+1,m-3)}
{\sqrt{(n+1)\,(n+2)}}~\betab^1\wedge\betab^2~\sum_{q\in\rel_{n+1}}\,
q\,\big|
{\stackrel{\scriptstyle n+1}{\scriptstyle q}}\,,\,m+3\big\rangle
\big\langle{\stackrel{\scriptstyle n+1}{\scriptstyle q}}\,,\,m-3
\big|
\label{betaprodholrel}\eeq
while in (\ref{curvnonholrelsCklrep}) we use
\bea
\betab^+_{n,m}\wedge\betab^-_{n,m}{}^\dag&=&-\,
\frac{\Lambda_{k,l}^+(n,m)\,\Lambda_{k,l}^-(n,m)}{2(n+1)} 
\label{betaprodnonholrel}\\ 
&& \times\,\sum_{q\in\rel_{n+1}}\,\Big[\sqrt{(n+1)^2-q^2}\,
\big(\beta^1\wedge\betab^1+\beta^2\wedge\betab^2\big)~\big|
{\stackrel{\scriptstyle n+1}{\scriptstyle q}}\,,\,m+3\big\rangle
\big\langle{\stackrel{\scriptstyle n-1}{\scriptstyle q}}\,,\,m+3
\big|\nonumber\\ && \hspace{2cm} +\,\sqrt{(n-q)^2-1}~
\beta^1\wedge\betab^2~\big|{\stackrel{\scriptstyle n+1}
{\scriptstyle q}}\,,\,m+3\big\rangle
\big\langle{\stackrel{\scriptstyle n-1}
{\scriptstyle q+2}}\,,\,m+3
\big| \nonumber\\ && \hspace{2cm} +\,\sqrt{(n+q)^2-1}~
\beta^2\wedge\betab^1~\big|{\stackrel{\scriptstyle n+1}
{\scriptstyle q}}\,,\,m+3\big\rangle
\big\langle{\stackrel{\scriptstyle n-1}
{\scriptstyle q-2}}\,,\,m+3\big|\,\Big] \ . \nonumber
\eea

Using (\ref{betaprodCkl1})--(\ref{betaprodnonholrel}) together with
\beq
\sum_{q\in\rel_n}\,q\=0 \qquad \mbox{and} \qquad
\sum_{q\in\rel_n}\,q^2\=\frac13\,n\,(n+1)\,(n+2) \ ,
\label{qsumids}\eeq
one can derive a number of trace identities which are
useful for deriving the dimensionally reduced gauge theory actions of
\S\ref{Eqgaugeth}. One has
\bea
\Tr\Big(\,\frac {\bar\b^\pm_{n,m}{}^\dag \wedge\cp2star
  \bar\b^\pm_{n,m}}
{\Lambda^\pm_{k,l}(n,m)^2} \,\Big)&=&2\pi^2\,(n+1\pm1)~\beta_{\rm vol}
\ , \nonumber \\[4pt]
\Tr\Big(\,\frac {\bar\b^\pm_{n,m}{}^\dag \wedge\bar\b^\pm_{n,m}\wedge
  \cp2star
\bigl(\bar\b^\pm_{n,m}{}^\dag \wedge\bar\b^\pm_{n,m}\bigr)^\dag}
{\Lambda^\pm_{k,l}(n,m)^4} \,\Big)&=&2\pi^2\,(n+1\pm 1)~\beta_{\rm
  vol} \ , \nonumber \\[4pt]
\Tr\Big(\,\frac { \bar\b^\pm_{n,m}\wedge\bar\b^\pm_{n,m}{}^\dag \wedge
  \cp2star  
\bigl(\bar\b^\pm_{n,m}\wedge\bar\b^\pm_{n,m}{}^\dag \bigr)^\dag}
{\Lambda^\pm_{k,l}(n,m)^4} \,\Big)&=&2\pi^2\,\frac {(n+1)^2}{n+1\pm
  1}~\beta_{\rm vol} \ , \\[4pt]
\Tr\Big(\,\frac { \bar\b^+_{n,m}\wedge\bar\b^-_{n+1,m-3} \wedge
  \cp2star\bigl(\bar\b^+_{n,m}\wedge \bar\b^-_{n+1,m-3} \bigr)^\dag}
{\Lambda^+_{k,l}(n,m)^2\,\Lambda^-_{k,l}(n+1,m-3)^2}
\,\Big)&=&2\pi^2\,\frac {(n+3)}{3}~\beta_{\rm vol} \ , \nonumber
\\[4pt]
\Tr\Big(\,\frac { \bar\b^+_{n,m}\wedge\bar\b^-_{n,m}{}^\dag \wedge
  \cp2star\bigl(\bar\b^+_{n,m}\wedge\bar\b^-_{n,m}{}^\dag \bigr)^\dag}
{\Lambda^+_{k,l}(n,m)^2\,\Lambda^-_{k,l}(n,m)^2} \,\Big)
&=&2\pi^2\,\frac {n\,(n+2)}{n+1}~\beta_{\rm vol} \ , \nonumber 
\eea
where $\Tr$ is the trace over $\SU(2)$ representations and $\cp2star$
is the Hodge duality operator on $\CP^2$ corresponding to the metric
(\ref{metric3}) with
\beq
\bar \b^1\wedge \cp2star \b^1 \= \bar \b^2\wedge \cp2star \b^2 \=
\b^1\wedge \cp2star \bar \b^1 \= \b^2\wedge \cp2star \bar\b^2 \= 
2\pi^2~\beta_{\rm vol} \ .
\eeq
Note that $\bar \b^1\wedge \cp2star \bar \b^1 =\bar \b^2\wedge
\cp2star \bar \b^2  =\b^1\wedge \cp2star  \b^2=\b^1\wedge \cp2star
\bar\b^2=0$, together with their hermitean conjugate equations.

\bigskip

\section{Index theorem on $\CP^2$\label{ASItheorem}}

\noindent
Spinors cannot be globally defined on $\CP^2$ due to a topological
obstruction.
However, globally well-defined spinors can be constructed by twisting
the Dirac operator on $\CP^2$ with half-integer powers $\Lcal_\mt$,
$\mt\in \Z + \frac 12$ of the monopole line bundle $\Lcal$. The index
of the Dirac operator associated with 
this twisted complex is computed by the Atiyah-Singer index theorem to
be~\cite{BDCN}
\begin{equation}
\nu_{\indint}\=\int_{\CP^2}\,\ch(\Lcal_\mt)\wedge\widehat A \= 
\mbox{$\frac12$}\,(\indint+1)\,(\indint+2)
\label{eq:rank1index}
\end{equation}
where $\ch(\Lcal)$ is the Chern character of $\Lcal$, $\widehat A$ is the
Atiyah-Hirzebruch class of $\CP^2$, and $\indint =\mt -\frac 32$ is an
integer.\footnote{The factor $-\frac 32$ here is essentially the power
  of $\Lcal$ arising from the $\urm(1)$ part of 
the holonomy in $\widehat A$. A factor of $3$ is the Euler
characteristic of $\CP^2$, and $-3$ is the first Chern number of the
canonical line bundle over $\CP^2$. The factor $-\frac 32$ arises
because, on a complex manifold,  
the spinor bundle involves the square 
root of the canonical line bundle. That this factor is not an integer
reflects the fact that the spinor bundle over $\CP^2$ does not exist
globally.} 
 
In the main text we use the index (\ref{nunb}) for higher rank
$\SU(3)$-equivariant bundles over $\CP^2$, and we will now derive this
formula here. From (\ref{QtoL}) the zero mode structure of the Dirac 
operator for spinor fields transforming under the holonomy group
$H=\SU(2)\times\urm(1)$, in the fundamental 
representation of $\SU(2)$ and in the background gauge field of
$\Qcal\otimes\Lcal_\mt$, is easily evaluated~\cite{BDCN}.  
Denoting this index index by 
$\nu_{\indint;1}$ we have, using (\ref{QtoL}), the formula
\begin{equation}
\nu_{\indint;1}\=\int_{\CP^2}\,\ch(\Qcal)\wedge\ch(\Lcal_\mt)\wedge
\widehat A \=3 \nu_{\indint} - \nu_{\indint-1}\=
(\indint+1)\,(\indint+3)
\label{eq:rank2index}
\end{equation}
where $\nu_b$ and $\nu_{b-1}$ have been evaluated with
(\ref{eq:rank1index}). The index with respect to all higher rank
bundles can be computed in terms of the 
rank one result (\ref{eq:rank1index}) by taking tensor powers of the
quotient bundle $\Qcal$, since
\begin{equation}
\int_{\CP^2}\,\ch(\Qcal^{\otimes n})\wedge\ch(\Lcal_\mt)\wedge
\widehat A
=\int_{\CP^2}\,\big(3-\ch(\Lcal_{-1})\big)^{\wedge n}\wedge
\ch(\Lcal_\mt)\wedge\widehat A \ .
\label{eq:ranknpreducibleindex}
\end{equation}
There is a technical issue, however, because $\Qcal^{\otimes n}$ is a
bundle of rank $2^n$ which is reducible in terms of $\SU(2)$
representations and it will be more convenient for our purposes to
decompose it into irreducible representations. 

The $n$-fold tensor product of the fundamental
representation of $\SU(2)\times\uo$ decomposes into irreducible
representations as 
\beq
\underline{(1,1)}^{\otimes n}
=\bigoplus_{t=0}^{\lfloor n/ 2 \rfloor}\; N_{t,n}\;
\underline{(n-2t,n)} \ , 
\label{nfoldproduct}
\eeq
where $N_{t,n}$ is the multiplicity
\beq
N_{t,n}=\frac{(n-2t+1)\,n!}{(n-t+1)!\,t!} \ .
\eeq
Consider the equivariant rank two instanton bundle $\Ical\to\CP^2$,
and its $n$-fold symmetric tensor product $\Ical_n$ given by
(\ref{SymnIcal}) which is an equivariant vector bundle over $\CP^2$ of
rank $n+1$. Its structure group is $\SU(2)$ and so it has no $\urm(1)$
charge. One then has 
\beq 
\Qcal^{\otimes n}=\Big(\,\bigoplus_{t=0}^{\lfloor n/ 2 \rfloor}\;
N_{t,n}\; \Ical_{n-2t}\,\Big)\otimes \Lcal_{n/2} \ .
\eeq

In \S\ref{InvSpinors} we use the index of the irreducible bundles
$\Ical_n\otimes \Lcal_{\ct+m/2}$ of rank $n+1$, with $\ct\in\Z+\frac
12$ a half-integer and $n\equiv m\mod 2$ so that $\underline{(n,m)}$
is a faithful representation of $\urm(2)$. With $\mt=\frac {m-n}2+\ct$
it is given by
\beq
\nu_{\indint;n}:=
\int_{\CP^2}\,\ch(\Ical_n)\wedge\ch(\Lcal_{\mt+n/2 })\wedge\widehat A
\label{eq:ranknpindex}
\eeq
rather than (\ref{eq:ranknpreducibleindex}). For given $n$ this can be 
calculated explicitly if we know all the lower $\nu_{\indint; n-2t}$
for $t\ge 1$, since the K-theory formula
\beq
\Ical_n=\big(\Qcal^{\otimes n}\otimes\Lcal_{-n/2}\big)\,\ominus\,
\Big(\,\bigoplus_{t=1}^{\lfloor n/ 2 \rfloor}\; N_{t,n}\;
\Ical_{n-2t}\,\Big)
\label{IcalnKtheory}\eeq
implies
\beq
\nu_{\indint; n}=
\int_{\CP^2}\,\ch(\Qcal^{\otimes n})\wedge\ch(\Lcal_\mt)\wedge
\widehat A -\sum_{t=1}^{\lfloor n/ 2 \rfloor}\, N_{t,n}
\,\nu_{\indint; n-2t} \ ,
\label{nuinduction}
\eeq
and the first term on the right-hand side of (\ref{nuinduction}) is
known explicitly from (\ref{eq:ranknpreducibleindex}) and
(\ref{eq:rank1index}). We already know $\nu_{\indint;0}=\nu_{\indint}$
from (\ref{eq:rank1index}) and $\nu_{\indint;1}$ from
(\ref{eq:rank2index}), so we now have all the necessary ingredients to
prove the formula (\ref{nunb}) by induction on $n$.

The index $\nu_{b;n}$ can be either positive or negative but its
magnitude always corresponds to the dimension (\ref{dimkl}) of some
irreducible representation of $\SU(3)$, as expected on general
grounds~\cite{BDHom}. For example, if $\indint\ge 0$ then the index
(\ref{nunb}) is the dimension of the $\SU(3)$-module $\underline
C^{n,\indint}$. Under the decomposition (\ref{fundrepdecomp}) the
irreducible $\SU(2)\times\urm(1)$ representation with largest monopole
charge is $\underline{(n,2\indint+ n)}$, where $\indint+\frac
n2=\mt+\frac n2 -\frac 32$ is the $\urm(1)$ charge of the bundle
$\Ical_n\otimes\Lcal_{\mt+n/2 }$ appearing in (\ref{eq:ranknpindex})
including the contribution $-\frac 32$ from the Atiyah-Hirzebruch
class $\widehat A$. We can represent this diagramatically using Young
tableaux, in the notation of (\ref{3to2tableau}). The Young diagram
for $\underline C^{n,\indint}$ is
\begin{equation}
\underbrace{\young(\ \Dots\ ,\  \Dots \ )}_{\indint}\kern -2.1pt
\raise 6.1pt\hbox{$\underbrace{\young(\ \Dots\ )}_n$} \ ,
\end{equation}
which gives the index $\nu_{b;n}$ when $\indint\ge 0$.  This contains
the irreducible $\SU(2)\times\urm(1)$ representation
\begin{equation}
\underbrace{\young(\times \Dots\times ,\times  \Dots \times
  )}_{\indint}\kern -2.0pt 
\raise 6pt\hbox{$\underbrace{\young(\times \Dots\times )}_n$}
\end{equation}
with $\urm(1)$ charge $\indint+\frac n2$, and this is the
representation content of (\ref{eq:ranknpindex}) when $\indint \ge0$.

The bundle $\Qcal^{\otimes n}$ appearing in (\ref{nuinduction}) has
monopole charge $\frac n2$, and when $n$ is odd the choice $\mt=-\frac
n2$ cancels this charge and corresponds to the pure $\SU(2)$ bundle
$\Ical_n$. Hence for odd $n$ taking $\indint=\mt-\frac 32 =
-\frac{n+3}2$ gives the index
\beq
\nu_{-\frac{n+3}2;\,n}=-\mbox{$\frac{1}{8}$}\,(n+1)^3 \ ,
\label{nun3}\eeq 
and corresponds to spinors coupling to pure anti-selfdual $\SU(2)$ 
gauge fields on $\CP^2$ in the $(n+1)$-dimensional irreducible
representation with no $\urm(1)$ component. Since $n=2k+1$ is odd this
is necessarily a spinor representation of $\SU(2)$, though the
magnitude of the index (\ref{nun3}) corresponds to the dimension of a
real representation $\underline{C}^{k,k}$ of $\SU(3)$. At the opposite
extreme, the integer (\ref{eq:rank1index}) is the index for spinors
coupling to a pure $\urm(1)$ self-dual gauge field on $\CP^2$ with no
$\SU(2)$ component, and $\nu_{\indint}$ equals the dimension of the
$\SU(3)$ representation $\underline{C}^{\indint,0}$ for $\indint \ge 0$ 
while $-\nu_{\indint}$ equals the dimension of
$\underline{C}^{0,|\indint|-3}$ for $\indint\le -3$.

\end{document}